\documentstyle[emulateapj]{article}

\begin{document}

\title{WHAT ARE NARROW-LINE SEYFERT 1 GALAXIES ?: TOWARD A 
       VIEWING-ANGLE-DEPENDENT UNIFIED MODEL FOR SEYFERT GALAXIES}

\author{Yoshiaki Taniguchi, Takashi Murayama, \& Tohru Nagao}

\affil{Astronomical Institute, Graduate School of Science, 
       Tohoku University, Aramaki, Aoba, Sendai 980-8578, Japan} 

\submitted{Submitted to the Astrophysical Journal}

\begin{abstract}
Narrow-line Seyfert 1 galaxies (NLS1s) have been recognized to 
comprise a distinct class of the Seyfert activity in addition to
type 1 and type 2 Seyfert galaxies (S1s and S2s, respectively); 
the NLS1s show strong optical Fe {\sc ii} emission lines as well as
high-ionization lines without optical broad emission lines.
Recently growing evidence has been accumulated that the 
NLS1s are viewed from nearly pole-on viewing angles.
However, there has been no theoretical model which explains
the following important correlations; 1)
weaker optical Fe {\sc ii} emitters always have wider emission-line widths
but stronger optical Fe {\sc ii} emitters have either wider or narrower 
line widths, and 2) S1s with wider line widths always have flatter
soft X-ray spectra but S1s with narrower line widths
have either steeper or flatter soft X-ray spectra.

Photoionization models predict that the partly-ionized zone
in the broad line region (BLR) with a disk-like configuration,
in which the optical Fe {\sc ii} emission
is thought to arise, is very optically thick;
e.g., the hydrogen column density is required to be as high as
$N_{\rm H} \gtrsim 10^{24.5}$ cm$^{-2}$.
Therefore, the visibility of the partly-ionized regions
is expected to be strongly viewing-angle dependent.
If we observe this disk-like BLR from an inclined viewing angle;
e.g., $i_{\rm view} \simeq  30^\circ$ [a typical viewing angle 
for broad-line S1s (hereafter BLS1s)], the Fe {\sc ii} emitting region
located in the far-side half disk cannot be seen entirely
because of the large optical depth while almost all the H$\beta$
emission can be seen because the ionized hydrogen is located in
the outer surfaces of BLR clouds. On the other hand,
if we observe the disk-like BLR from a nearly pole-on view,
we can see both Fe {\sc ii} and H$\beta$, resulting in a higher
Fe {\sc ii}/H$\beta$ ratio together with a narrower line width
with respect to those observed from an inclined viewing angle.
This explains the first important correlation.

Recent discovery of relativistic outflows in some NLS1s suggests 
that the nuclear radio jet interacts with the dense ambient gas
very close to the central engine. We show that 
this can be responsible for the formation of hot plasma with 
kinetic temperatures of $T_{\rm kin} \sim 10^6$ K,
giving rise to the production of soft X-ray photons.  
The black-body radiation from the hot plasma explains the observed
steeper photon indices in the soft X ray if
$T_{\rm kin} \simeq$ (1 -- 2) $\times 10^6$ K
and $\beta = v_{\rm jet} / c  \simeq$ 0.2 -- 0.7.
The kinetic Doppler effect increases the soft X-ray luminosities
if we observe S1s from a nearly pole-on view
given the above range of $\beta$. These explain why NLS1s
tend to have steeper soft X-ray spectra together with higher
soft X-ray luminosities. Since the pole-on view model for NLS1s
implies that BLS1s are viewed from intermediate orientations,
the brightening due to the kinematic Doppler effect is generally weak
for the BLS1s and thus the soft X-ray excess emission in the BLS1s
is less dominant than that in the NLS1s.
Furthermore, the extinction of soft X-ray photons 
due to dust grain above the dusty
tori are expected to more serious for the BLS1s than for the NLS1s.
We thus suggest that these orientation effects are responsible for 
the second important correlation. It is also understood why soft X-ray 
surveys tend to pick up NLS1s preferentially. 

It is known that there is a correlation between
the soft X-ray photon index ($\Gamma$)
and the Fe {\sc ii} $\lambda$4570/H$\beta$ ratio for both NLS1 
and BLS1s. We also newly find another correlation between
$\Gamma$ and the soft X-ray luminosity for the NLS1s studied by
Boller, Brandt, \& Fink. These correlations suggest that the 
strength of Fe {\sc ii} emission 
is intimately related to the soft X-ray emission.
Since the ultraviolet continuum emission from the optically-thick
accretion disk with $T \sim 10^5$ K 
cannot create partly-ionized regions in the disk-like BLR efficiently,
we suggest that soft X-ray photons from the jet-driven shocked regions
with $T \sim 10^6$ K 
are responsible for the formation of the Fe {\sc ii} emitting regions.
This provides a causal relationship between the strong Fe {\sc ii} 
emission and the excess soft X-ray emission.

Finally, we propose a viewing-angle-dependent unified
model for Seyfert nuclei;
1) $0^\circ \leq i_{\rm view} \lesssim 10^\circ$ for NLS1s,
2) $10^\circ \lesssim i_{\rm view} \lesssim 30^\circ$ for BLS1s,
3) $30^\circ \lesssim i_{\rm view} \lesssim 50^\circ$ for type 2 
Seyferts with the hidden BLR (S2$^+$s), and
4) $50^\circ \lesssim i_{\rm view} \leq 90^\circ$ for type 2 Seyferts
without  the hidden BLR (S2$^-$s) where $i_{\rm view}$ is the viewing
angle toward the BLR disk around the central engine.
Since the NLS1 phenomenon is also observed in radio-quiet quasars,
it is strongly suggested that the class of NLS1s
is the radio-quiet equivalent of the class of Blazers
in radio-loud active galactic nuclei.

\end{abstract}




\section{INTRODUCTION}

Seyfert nuclei are typical active galactic nuclei (AGN) in the
nearby universe and their central engine is thought to be gas accretion
onto a supermassive black hole (e.g., Rees 1984; Blandford 1990).
The Seyfert nuclei have been broadly classified into two types 
based on the presence or absence of broad emission-line region (BLR)
in their optical spectra (Khachikian \& Weedman 1974); Seyferts with 
BLR are type 1 (hereafter S1) while those without BLR are type 2 (S2).
These two types of Seyfert nuclei are now unified by introducing 
the viewing angle dependence toward the central engine given that
a geometrically and optically-thick,
dusty torus surrounds the central engine 
(Antonucci \& Miller 1985; see for a review Antonucci 1993).

In addition to typical S1s and S2s,
narrow-line Seyfert 1 galaxies (NLS1s) have also been recognized
to comprise a distinct class of Seyfert nuclei since Davidson \& Kinman 
(1978) noticed their particular observational properties
[Koski 1978; Phillips 1978; Osterbrock \& Pogge 1985 (hereafter OP85);
Halpern \& Oke 1987; van Groningen \& de Bruyn 1989;
Puchnarewicz et al. 1992, 1995a, 1995b; Mason, Puchnarewicz, \& Jones 1996].
It is known that NLS1s share about ten percent of S1s
in the hard X-ray selected sample (Stephens 1989). 
A large number of new NLS1s have been found in recent soft X-ray
surveys of AGN [Puchnarewicz et al. 1992, 1996; Walter \& Fink 1993; 
Boller, Brandt, \& Fink 1996 (hereafter BBF96); Xu, Wei, \& Hu 1999; 
Grupe et al. 1999; Edelson et al. 1999].
However, it is still not understood what the class of NLS1s is 
within the context of the current unified model. 
In order to construct a more perfect unified model
of Seyfert nuclei, it is necessary to understand
what NLS1s are and their relation to typical S1s and S2s.

In this paper, we give a summary of important observational
properties of NLS1s (section 2) and possible models for NLS1s
(section 3). Several observational tests are made in section 4.
Kinematical and statistical considerations are given in 
sections 5 and 6, respectively.
Then, in section 7,  we propose a new model for NLS1s, which is 
consistent with almost all the observational properties of NLS1s. 
Finally, a viewing-angle-dependent unified model for Seyfert nuclei
is presented in section 8. In order to avoid confusion,
we refer typical S1s as broad-line Seyfert 1s (BLS1s).
Therefore, it is meant that S1s = NLS1s + BLS1s.

\section{OBSERVATIONAL PROPERTIES OF NLS1s}

In this section, we give a summary of important observational properties
of NLS1s in the optical, ultraviolet, soft and hard X rays, radio,
and infrared. The individual items are labeled; e.g., [O1] (the first 
item in the optical), [O2] (the second item in the optical), and so on.
We will refer the observational properties using these item codes
in later discussion.

\subsection{Optical}

\begin{description}

\item{[O1]} Balmer emission lines are slightly broader 
than forbidden lines such as 
[O {\sc iii}]$\lambda$5007 emission but their line widths (FWHMs) 
are typically $\simeq$ 1000 
km s$^{-1}$  (Davidson \& Kinman 1978; 
OP85; see also Mason, Puchnarewicz, \& Jones 1996). 
The narrowest FWHM is $\approx$ 300 km s$^{-1}$ 
(Davidson \& Kinman 1978; OP85)
while the widest one amounts
to $\approx$ 2000 km s$^{-1}$ (Vaughan et al. 1999).

\item{[O2]} The profiles of BLR emission of NLS1s are generally different
from those of BLS1; i.e., a normal broad-line profile has a more 
dominant core than a NLS1 profile (Robinson 1995).

\item{[O3]} The equivalent widths of Blamer emission
lines are on average lower than
those of BLS1s (OP85; Goodrich 1989); 
e.g., EW(H$\beta$) $\simeq 32 \pm 16$ \AA~ for the NLS1s and
EW(H$\beta$) $\simeq 88 \pm 37$ \AA~ for the BLS1s (OP85).

\item{[O4]} The time variability of 
Balmer emission lines is similar to that observed
in BLS1s (Giannuzzo \& Stripe 1996). However, there is a marginal
trend that the variability in NLS1s is weaker than that in BLS1s
(Giannuzzo et al. 1998).

\item{[O5]} The [O {\sc iii}]$\lambda$5007/H$\beta$ ratio is smaller 
than 3 (OP85; Goodrich 1989).

\item{[O6]} High-ionization emission lines like [Fe {\sc vii}]$\lambda$6087 
and [Fe {\sc x}]$\lambda$6735 are present 
(Davidson \& Kinman 1978; OP85; Nagao et al. 1999b).
The [Fe {\sc vii}]$\lambda$6087/[O {\sc iii}]$\lambda$5007 ratio of
NLS1s is on average similar to that of BLS1s
(Nagao, Murayama, \& Taniguchi 1999a).

\item{[O7]} Optical Fe {\sc ii} emission lines are generally stronger 
than those of BLS1s. There is a significant correlation between 
the Fe {\sc ii} $\lambda$4570/H$\beta$
intensity ratio and the FWHM of H$\beta$ emission (Zheng \& O'Brien 1990;
Marziani et al. 1996);
weak Fe {\sc ii} emitters always have larger FWHMs but stronger Fe {\sc ii}
emitters have either larger or smaller FWHMs.

\item{[O8]} Another correlation is found between 
Fe {\sc ii} $\lambda$4570/H$\beta$ and
[O {\sc iii}]$\lambda$5007/H$\beta$; Fe {\sc ii}/H$\beta$
tends to be stronger with decreasing 
[O {\sc iii}]/H$\beta$ (Boroson \& Green 1992).

\item{[O9]} Spectropolarimetry of 17 NLS1s 
shows that no NLS1 has hidden BLR and the polarization
detected in six NLS1s are attributed to the dust scattering
rather than the electron scattering (Goodrich 1989); note that 
the polarization observed in most S2s is attributed to the
electron scattering (Antonucci \& Miller 1985; Miller \& Goodrich
1990; Tran 1995).

\item{[O10]} A summary of the emission-line components observed in
NLS1s, BLS1s, and S2s is given in Table 1 (see also [U2]).

\end{description}

\begin{deluxetable}{lcccccc}
\tablenum{1}
\tablewidth{0em}
\tablecaption{UV/optical emission-line
components observed in BLS1s, NLS1s, and S2s}
\tablehead{
 & \colhead{Component} & \colhead{HI-BLR\tablenotemark{a}}
 & \colhead{LI-BLR\tablenotemark{b}} & \colhead{NLR} 
 & \colhead{HINER} & \colhead{Fe {\sc ii}} \nl
}
\startdata
BLS1s &  &  Yes  & Yes & Yes & Yes     & Yes \nl
NLS1s &  &  Yes  &  No & Yes & Yes     & Yes \nl
S2s   &  &  No   &  No & Yes & Yes/No  & No  \nl
\enddata
\tablenotetext{a}{High-ionization BLR}
\tablenotetext{b}{Low-ionization BLR}
\end{deluxetable}

\subsection{Ultraviolet}

\begin{description}

\item{[U1]} Ultraviolet (UV) emission-line ratios 
(e.g., C {\sc iv}$\lambda$1549/Ly$\alpha$)
of NLS1s are similar to those of BLS1s (Crenshaw et al. 1991).

\item{[U2]} Blue wing emission is present in UV emission lines, 
Ly$\alpha$ and C {\sc iv}$\lambda$1549. 
Their line widths amount to $\approx$ 5000 km s$^{-1}$,
being comparable to those of BLS1s
(Rodr\'iguez-Pascual, Mas-Hesse, \& Santos-Lle\'o 1997).

\item{[U3]} The UV luminosities of NLS1s tend to 
be less luminous than those of BLS1s (Rodr\'iguez-Pascual et al. 1997).

\end{description}

\subsection{X ray}

\begin{description}

\item{[X1]} NLS1s tend to be stronger soft X-ray emitters with respect to
BLS1s and  often have ROSAT photon indices steeper than $\Gamma_{\rm soft} = 3$
where the energy range is between 0.1 keV and 2.4 keV
(Puchnarewicz et al. 1992; BBF96; Wang, Brinkmann, \& Bergeron 1996;
see for earlier indications Wilkes, Elvis, \& McHardy 1987;
Zheng \& O'Brien 1990). The average photon index for the NLS1s is
$\overline{\Gamma}_{\rm soft} \simeq 3.13$ while that for the BLS1s is 
$\overline{\Gamma}_{\rm soft} \simeq 2.34$ (BBF96)

\item{[X2]} There is a significant correlation between the soft X-ray spectral 
index and the FWHM of H$\beta$ emission (Joly 1991; Zheng \& O'Brien 1990);
S1s with larger FWHM(H$\beta$) (i.e., BLS1s) always have flat spectra but 
S1s with smaller FWHM(H$\beta$) (i.e., NLS1s) 
emitters have either steep or flat spectra (BBF96;
see also Lawrence et al. 1997).

\item{[X3]} Hard X-ray spectra of NLS1s with steeper soft X-ray spectra
are also steeper than those of BLS1s; 22 NLS1s studied with ASCA
have 2 -- 10 keV photon indices of $\Gamma_{\rm hard} \simeq$
1.6 -- 2.5 with a mean value of $\sim 2.1$
(Brandt, Mathur, \& Elvis 1997; Leighly 1999b;  Vaughan et al. 1999).
Note that BLS1s have $\overline{\Gamma}_{\rm hard} \approx$ 1.9
(Nandra \& Pounds 1994; Nandra et al. 1997).

\item{[X4]} Broad  absorption features centered 
in the energy range 1.1 -- 1.4 keV
or so-called warm absorption is found in $\approx$ 40\% (i.e., 9/22) 
of NLS1s (Leighly et al. 1997; Vaughan et al. 1999; Leighly 1999b). 
All these have narrower H$\beta$ line widths; FWHM $\lesssim$ 1000 km s$^{-1}$
(Vaughan et al. 1999). However, it is noted that the detection rate of 
so-called warm absorbers in NLS1s appears lower than that in BLS1s
(Wang et al. 1996; Leighly 1999b).

\item{[X5]} Time variability in the soft X ray is often observed; the
timescale is shorter than one day (BBF96).
This corresponds to a size less than one light day, being 
smaller than the characteristic size of the BLR of BLS1s.

\item{[X6]} Time variability is also observed in the hard X ray (Leighly 1999a).
According to the time series analysis based on the ASCA data of
24 NLS1s, the excess variance from the NLS1s light curves is
inversely correlated with the X-ray luminosities.
Furthermore, the excess variance for the NLS1s is flatter than that for 
BLS1s. 
The amplitude of the variability in the hard X ray is correlated
with  the strength of the soft excess; the NLS1s with
stronger soft-excess
tend to show the larger amplitude variability (Leighly 1999b).

\item{[X7]} Soft X-ray observations show that there is no evidence for
large neutral hydrogen column densities over the Galactic column,
suggesting that the obscuration by dust grains is unimportant (BBF96).

\item{[X8]} There is a positive correlation between  $\Gamma_{\rm soft}$
and the UV (1375 \AA)-to-X ray (2 keV) flux ratio for S1s and NLS1s
follow this correlation (Walter \& Fink 1993).

\end{description}

\subsection{Radio}

\begin{description}

\item{[R1]} The radio continuum luminosities of NLS1s lie in 
the range found for BLS1s (Ulvestad, Antonucci, \& Goodrich 1995).
However, as noted by them, their sample is 
not a statistically complete sample.

\item{[R2]} The median radio size is no larger than 300 pc;
i.e., most objects are not resolved by VLA observations
(Ulvestad et al. 1995).
Since the median radio size of BLS1s is 350 pc (Ulvestad \& Wilson 1989),
the NLS1s have radio sources with sizes similar to those in the BLS1s
[here a Hubble constant $H_0$ = 75 km s$^{-1}$ Mpc$^{-1}$ is adopted] .

\item{[R3]} Among the seventeen NLS1s, only three have measurable
radio axes; two NLS1s (Mrk 766 and Mrk 1126)
have radio major axes perpendicular to the optical polarization
while the remaining one (Mrk 957) has a radio major axis
parallel to the optical polarization
(Ulvestad et al. 1995).
 
\end{description}

\subsection{Infrared}

\begin{description}

\item{[I1]} Mid- and far-infrared luminosities based on IRAS observations
(i.e., 12 $\mu$m, 25 $\mu$m, 60 $\mu$m, and 100 $\mu$m)
of NLS1s are similar to those of BLS1s 
(Rodr\'iguez-Pascual et al. 1997).

\item{[I2]} The $L$-band (3.5 $\mu$m)-to-IRAS 25 $\mu$m flux ratio,
which is sensitive to the orientation of dusty tori, of NLS1s
is on average similar to that of BLS1s (Murayama et al. 1999; see also
Murayama, Mouri, \& Taniguchi 2000).

\end{description}

\section{A SUMMARY OF PREVIOUS INTERPRETATIONS}

Possible interpretations of NLS1s discussed previously are 
summarized below (see for good reviews BBF96;
Giannuzzo \& Stripe 1996).
Note that five among the following seven models (I, II, III, V,
and VI) assume that the BLR line width is dominated by the rotational
motion around a central compact object.

\begin{description}

\item{I)} The pole-on view model (OP85; Goodrich 1989; Stephens 1989;
Puchnarewicz et al. 1992):
NLS1s are basically BLS1s but are viewed from a more face-on view
(i.e., nearly pole-on view). This model requires that 
the BLR has a disk-like configuration. 

\item{II)} The low-$M_\bullet$ model (Ross \& Fabian 1993): 
NLS1s are basically BLS1s but their narrow line widths are 
attributed to the lower masses of the central black holes;
e.g., $M_\bullet \sim 10^6 M_\odot$ for NLS1s while
$M_\bullet \sim 10^7 M_\odot$ for BLS1s.

\item{III)} The distant-BLR model (Mason et al. 1996):
NLS1s are basically BLS1s but their narrow line widths are attributed to
that the BLR in NLS1s
is located at $r_{\rm BLR} \sim$ 0.1 pc -- 1 pc.
Note that 
a typical radial distance of the BLR in BLS1s is 
$\sim$ 0.01 pc (e.g., Peterson 1993). 
A theoretical consideration for this model was given by
Wandel \& Boller (1998).

\item{IV)} The no-BLR model (e.g., Giannuzzo \& Stripe 1996):
NLS1s are basically BLS1s but there is little gas in the BLR.

\item{V)} The partly-hidden-BLR model (e.g., Giannuzzo \& Stripe 1996):
NLS1s are basically BLS1s but NLS1s are objects seen at relatively
large inclination angles and thus only outer parts of the  BLR
can be seen. Since the outer parts of the BLR are seen in NLS1s, 
NLS1s are observed to be different from S2s whose BLR is
completely hidden by dusty tori.

\item{VI)} The intermediate-zone model (Mason et al. 1996):
An intermediate-velocity (FWHM $\simeq$ 1000 km s$^{-1}$) line-emitting region 
produces significant amounts of the permitted and forbidden emission line fluxes
in NLS1s. Since the estimated radial distance of this intermediate zone is 
$\sim$ 1 pc, this region is expected to be located between the BLR and the NLR
(narrow-line region).
The absence of the BLR in NLS1s is not understood solely by this model unless the 
intermediate zone is spatially identical to the distant BLR (see Model III).

\item{VII)} A supermassive analog of Galactic black hole candidates
(hereafter the GBHC model; Pounds et al. 1995):  
NLS1s are the supermassive black hole analogs of Galactic 
black hole candidates in the high state which are thought to be
accreting at a larger fraction of the Eddington limit than 
those in the low state (e.g., Tanaka 1990).

\end{description}

\section{OBSERVATIONAL TESTS}

As summarized in section 3, there are seven possible models for the NLS1s.
Since there are many interesting observational properties of the NLS1s 
as summarized in section 2, we can make various observational tests to
reject some models. 

\subsection{Evidence Against the Distant-BLR Model}

The distant-BLR model requires that the radial distance of the BLR is
more distant than the typical value for BLS1s; i.e., $\sim 0.01$ pc
(e.g., Peterson 1993).
If we adopt a radial distance of the BLR of NLS1s $r_{\rm BLR} = 0.1$ pc, 
we obtain
a maximum line width FWHM$^0$(BLR) $\approx 2 \times v_{\rm rot}
\simeq 1320 M_{\bullet, 7}^{1/2} r_{\rm BLR, 0.1}^{-1/2}$ km s$^{-1}$
where $v_{\rm rot} = (G M_\bullet / r_{\rm BLR})^{1/2}$ is the Keplerian
rotational velocity and $M_{\bullet, 7}$ is the
mass of the supermassive black hole in units of $10^7 M_\odot$.
Thus this model explains the narrow line widths of the BLR emission.
However, Robinson (1995) found that the profiles of BLR emission of 
NLS1s are generally different
from those of BLS1; i.e., a normal broad-line profile has a more
dominant core than a NLS1 profile [O2].
This property cannot be explained by the distant-BLR model.

The optical monitoring observations of NLS1s
promoted by Giannuzzo and coworkers
have shown that there is no significant difference between the 
optical variability properties of NLS1s and BLS1s [O4].
This provides also evidence against the distant-BLR model.

Recently, Wandel \& Boller (1998) proposed a theoretical basis
for the distant-BLR model;
a stronger photoinizing continuum present in NLS1s
can be responsible for the ionization of BLR clouds
at larger radii where the Keplerian velocities are lower.
However, in their model, it is not well understood
why only NLS1s have such stronger-photoinizing  continua.
Therefore, in order to accept this model, one will have to
explain the origin of the stronger continuum emission.
Furthermore, 
if we adopt this model, the number ratio between NLS1s and BLS1s
can be attributed to a variety of the radial distance of BLR in Seyfert nuclei.
There may be many parameters to determine $r_{\rm BLR}$ in each Seyfert;
the dynamical stability of a rotating gaseous disk (or a ring) around
a SMBH, the angular momentum of the BLR gas, the phase of gas accretion,
and so on. Therefore, at present, this model may not give a self-consistent
explanation of NLS1s.

\subsection{Evidence Against the No-BLR Model}

It has been considered that the intense soft X-ray emission
can prevent the formation of the BLR clouds close
to the central engine (Guilbelt, Fabian, \& McCray 1983;
White, Fabian, \& Mushotzky 1984).
However, although there is no evidence for the BLR in the optical spectra
of NLS1s,
Rodr\'iguez-Pascual et al. (1997) found the blue
wing emission in UV emission lines, Ly$\alpha$ and
C {\sc iv}$\lambda$1549 [U2]. Since the line widths of the blue wing emission
amount to $\approx$ 5000 km s$^{-1}$, 
it turns out that the NLS1s observed by them have the BLR.
Accordingly, we can reject the no-BLR model.

However, one problem remains. Why do only UV lines show evidence
for the BLR while there is no evidence for the BLR in the optical ?
Since the UV BLR shows the blue wing emission, it is likely that 
the UV BLR is associated with some outflow activity. 
This means that there are two kinds of BLR; one is the disk-like
BLR which emits optical Blamer lines and the other is the jet-like
BLR which emits UV BLR emission lines like Ly$\alpha$ and C {\sc iv}
(see section 7.2 for more detail).
This picture appears consistent with the recent analysis of UV and 
optical emission lines of AGN (Sulentic et al. 1995;
Marziani et al. 1996; see also Dultzin-Hacyan, Taniguchi, \& 
Uranga 1999; Taniguchi, Dultzin-Hacyan, \& Murayama 1999).

\subsection{Evidence Against the Intermediate-Zone Model}

Mason et al. (1996) found the 
intermediate-velocity (FWHM $\simeq$ 1000 km s$^{-1}$) line-emitting 
region which produces significant amounts of the permitted and 
forbidden emission line fluxes in one of NLS1s, RE J1034+396.
However, the estimated radial distance of this intermediate 
zone,  $\sim$ 1 pc, is similar to that of the inner wall of dusty tori 
(e.g., Taniguchi \& Murayama 1998 and references therein).
If the line width of the rotational motion of the dusty tori,
we obtain FWHM $\simeq 1320 
M_{\bullet, 8}^{1/2} r_1^{-1/2}$ km s$^{-1}$
$\simeq 1320 M_{\bullet, 7}^{1/2} r_{0.1}^{-1/2}$ km s$^{-1}$ 
where $M_{\bullet, 8}$ and $M_{\bullet, 7}$ are the 
black hole mass in units of $10^8 M_\odot$ and $10^7 M_\odot$, 
respectively, and $r_1$ and $r_{0.1}$ are the radial distance of 
the inner wall in units of 1 pc and 0.1 pc, respectively. 
Since the inner wall of dusty tori is one of the important 
emission-line regions (Pier \& Voit 1995; Murayama \& Taniguchi 1998a, 
1998b;  Kramer et al. 1998), the modest interpretation 
for the intermediate zone found by Mason et al. (1996) is that such 
intermediate-velocity (i.e., FWHM $\simeq$ 1000 km s$^{-1}$) lines
arise from the inner wall of dusty tori (cf. Sulentic \& Marziani
1998).
Therefore, the presence of such intermediate-velocity
line-emitting regions may not be an essentially important property
of the NLS1s.

\subsection{Evidence Against the Partly-Hidden-BLR Model}

The partly-hidden-BLR model means that
NLS1s are objects seen at relatively
large inclination angles and thus only the outer parts of the  BLR
can be seen; i.e., NLS1s could be regarded as an intermediate class between
BLS1s and S2s. Since this model allows the presence of inner parts of the BLR,
this is different from the distant-BLR model.

However, if this is the case, the optical spectropolarimetry could detect
hidden BLRs in the polarized spectra of NLS1s.
However, Goodrich (1989) could not find any evidence for the hidden BLR [O9].
Although he found the polarized continuum emission in six NLS1s among 18,
the polarization is attributed to the scattering by dust grains rather than
free electrons. Dust grains which produce the observed polarization
may be located either in the NLR (e.g., Netzer \& Laor 1993) or
in the inner wall of dusty tori.
Since the polarization of S2s is attributed to the electron
scattering (Antonucci \& Miller 1985; Miller \& Goodrich 1990; Tran 1995),
it is unlikely that NLS1s are viewed from nearly the same viewing angles
as those for S2s. 

It is known that the inner wall of dusty tori produces  
the high-ionization emission lines such as [Fe {\sc vii}] and [Fe {\sc x}]
because this region has a larger ionization parameter and relatively higher
electron densities, e.g., $n_{\rm e} \sim 10^{7-8}$ cm$^{-3}$
(e.g., Pier \& Voit 1995; Murayama \& Taniguchi 1998b).
Murayama \& Taniguchi (1998a) found that S1s have excess
[\ion{Fe}{7}] $\lambda$6087 emission with respect to S2s and
proposed that the high-ionization
nuclear emission-line region (HINER: Binette 1985; Murayama, Taniguchi,
\& Iwasawa 1998) traced by the
[\ion{Fe}{7}] $\lambda$6087 emission resides in the inner wall of
dusty tori. Murayama \& Taniguchi (1998b)
constructed new dual-component (i.e.,
a typical NLR with a HINER torus) photoionization models
and showed that the observations are well explained if the
torus emission contributes to $\sim 10$ \% of the NLR emission. 
Therefore, the significant excess emission of such high-ionization lines
in BLS1s can be attributed to the relative importance of the torus HINER
with respect to S2s. Recently, Nagao et al. (1999a) found that the strength
of [Fe {\sc vii}]$\lambda$6087 emission relative to [O {\sc iii}]$\lambda$5007
of NLS1s is similar on average to that of BLS1s.
Therefore, the viewing angles to the dusty tori for NLS1s are not different
significantly from those for BLS1s.

Recently Murayama et al. (1999) made a mid-infrared test for NLS1s.
They compared the $L$-band (3.5 $\mu$m)-to-IRAS 25 $\mu$ flux ratio,
which is sensitive to the orientation of dusty tori, among NLS1s, BLS1s, and S2s
and found that the ratio of NLS1s
is on average similar to that of BLS1s.
This suggests again that the average viewing angles toward the NLS1s are
not different significantly from those toward the BLS1s.

All the above observations appear to be inconsistent with the partly-hidden-BLR
model. Thus we can reject this model.

\section{KINEMATICAL CONSIDERATION}

\subsection{Introduction}

Among the seven models, we have rejected the four models;
the distant-BLR model,  the no-BLR model,
the intermediate-zone model, and the partly-hidden-BLR model.
We also do not consider 
the GBHC model in later discussion although we will give some
comments on this model in section 7.4.
Now the following two models remain; 
the pole-on view model, and the low-$M_\bullet$ model.
Prior to going to construct a possible model for NLS1s, we investigate 
these two models from kinematical points of view.

A typical FWHM of the BLR of BLS1s is 5000 -- 10000  km s$^{-1}$ 
(e.g., Osterbrock 1989; Peterson 1997; Eracleous \& Halpern 1994).
Although double-peaked BLR profiles with FWHM $\sim$ 12000 km s$^{-1}$
are also found in a large number of AGN,
they are more often found in radio-loud AGN (Steiner 1981; Eracleous
\& Halpern 1994).
Since our main purpose is to investigate the origin of NLS1s (i.e., radio-quiet
AGN), we adopt FWHM(BLR) = 6000 km s$^{-1}$ for BLS1s in this discussion
(Eracleous \& Halpern 1994).

Recent reverberation mapping for a large number of BLS1s has shown that 
the line widths of the BLR are dominated by the rotational motion
(e.g., Peterson 1993; Wanders et al. 1995). 
One important aspect is that there are two alternative options for the
geometrical properties of the BLRs; a disk-like configuration or a jet-like one.
Recently Rodr\'iguez-Pascual et al. (1997) found 
the blue wing emission in UV emission lines, Ly$\alpha$ and
C {\sc iv}$\lambda$1549 in some NLS1s [U2], implying the presence of the
jet-like BLR. Even though, these NLS1s have narrower optical BLR
emission lines (i.e., H$\alpha$,  H$\beta$, and Fe {\sc ii}) whose FWHMs are
$\approx 1000$ km s$^{-1}$. It seems reasonable to assume that 
these narrower optical BLR lines arise from the disk-like BLR.

\subsection{An Intrinsic FWHM of the Disk-like BLR}

First let us estimate an intrinsic FWHM of the BLR in BLS1s.
Although we adopt FWHM(BLR) = 6000 km s$^{-1}$ for BLS1s, 
this is not an intrinsic one because we do not always observe the BLR disk
from a perfect face-on view. In fact, the MIR test [I2] suggests
that the viewing angle toward the dusty tori of BLS1s lies
in a range between $0^\circ$ and $30^\circ$ -- $45^\circ$
(Murayama et al. 1999, 2000).
Although either the BLR disk or the dusty torus or both may be warped
more or less (Pringle 1997; Nishiura, Murayama, \& Taniguchi 1998),
it is likely that we observe the BLR disk in BLS1s from nearly the same 
viewing angles estimated above for the dusty tori.
It is also reported that the inner accretion disk probed by Fe K line
in the hard X ray is inclined by $\simeq 30^\circ$ to the lines of sight
for a large number of BLS1s (e.g., Tanaka et al. 1995; Nandra et al. 1997).
Therefore, it seems reasonable to assume that an average viewing angle to
BLS1s is $\overline{i}_{\rm view} \approx 30^\circ$.
Then we obtain an intrinsic FWHM of the BLR in BLS1s,

\begin{equation}
{\rm FWHM}^0({\rm BLR}) =
{{\rm FWHM(BLR)} \over {\rm sin} ~\overline{i}_{\rm view}}
= 12000 ~~ {\rm km ~ s}^{-1}. 
\end{equation}
It is noted that this intrinsic FWHM is almost comparable to a
typical FWHM of the double-peaked BLR (Eracleous \& Halpern 1994).

We examine whether or not the simple Keplerian rotation is 
responsible for this line width. If we adopt
$M_\bullet = 10^7 M_\odot$ and $r_{\rm BLR} = 0.01$ pc,
we obtain FWHM$^0$(BLR) = 2 $v_{\rm rot} \simeq 4200 M_{\bullet, 7}^{1/2} 
r_{\rm BLR, 0.01}^{-1/2}$ km s$^{-1}$ where $M_{\bullet, 7}$ is the black hole
mass in units of $10^7 M_\odot$ and $r_{\rm BLR, 0.01}$ is the radial distance
of the BLR from the galactic nucleus in units of 0.01 pc.
The mass of supermassive black holes in BLS1s lies in a range between
$10^7 M_\odot$ and $10^8 M_\odot$ (e.g., Miyoshi et al. 1995;
Greenhill  et al. 1996;  Nishiura \& 
Taniguchi 1998 and references therein).
If we adopt $M_\bullet = 10^8 M_\odot$, we obtain

\begin{equation}
{\rm FWHM}^0({\rm BLR}) \simeq 13200 M_{\bullet, 8}^{1/2} 
r_{\rm BLR, 0.01}^{-1/2} ~~ {\rm km ~ s}^{-1}.
\end{equation}
The following combinations also give the same FWHM$^0$(BLR);
($M_\bullet$, $r_{\rm BLR}$) = ($10^7 M_\odot$, 0.001 pc), and
($3 \times 10^7 M_\odot$, 0.003 pc).
Therefore, the intrinsic FWHM for the BLR appears to be explained
by the disk rotation without invoking other motions.

\subsection{Origin of the Narrow FWHM of NLS1s}

Next we consider why FWHM(BLR) $\simeq 1000$ km s$^{-1}$ for NLS1s
for the following two cases: 1) the pole-on view model, and
2) the low-$M_\bullet$ model.

1) The pole-on view model: 
This model implies that the observed narrow FWHMs for
NLS1s are attributed to smaller viewing angles toward the BLR.
Therefore, given the intrinsic FWHM of the BLR in S1s
FWHM$^0$(BLR) $\simeq$ 12000 km s$^{-1}$ with the typical FWHM(BLR)
$\simeq$ 1000 km s$^{-1}$ for the NLS1s, we obtain
a critical viewing angle toward NLS1s  $i_{\rm cr, NLS1} \simeq$
sin$^{-1}$ (1000 km s$^{-1}$/12000 km s$^{-1}$) $\simeq 4\fdg8$; i.e.,
$0^\circ \leq i_{\rm view} \leq i_{\rm cr, NLS1}$ for NLS1s.
On the other hand, if we adopt the maximum value of
FWHM(BLR) $\simeq$ 2000 km s$^{-1}$ for NLS1s (Vaughan et al. 1999),
we obtain $i_{\rm cr, NLS1} \simeq 9\fdg6$.
This seems to be a more appropriate estimate.

2) The low-$M_\bullet$ model: If we adopt $M_\bullet = 10^6 M_\odot$,
we obtain FWHM$^0$(BLR) $\simeq 1320 M_{\bullet, 6}^{1/2} 
r_{\rm BLR, 0.01}^{-1/2}$ km s$^{-1}$.
Since the optical reverberation mapping shows that the shortest
variability timescale is about one week, i.e., corresponding to a linear 
dimension $\sim$ 0.01 pc (e.g., Peterson 1993), the mass of SMBHs 
is required to be less massive than $10^6 M_\odot$.
The dynamical mass of nuclei has been estimated for
a number of BLS1s (Wandel \& Yahil 1985; Padovani \& Rafanelli 1988;
Padovani, Burg, \& Edelson 1989;
Padovani 1989; Koratkar \& Gaskell 1991) and for S2s with the hidden BLR
(Nishiura \& Taniguchi 1998). However, there is no dynamical estimate
for the nuclei of NLS1s.
Recently Hayashida et al. (1998; see also Hayashida 1997) proposed a
new method to estimate the central black hole masses based on
the X-ray flux variability. They obtained $M_\bullet \simeq 4.93
\times 10^4 M_\odot$ for one of NLS1s, NGC 4051. This mass is
smaller  by two or three orders of magnitude than the typical mass of
Seyfert nuclei. Indeed, the recent reverberation mapping gives
$M_\bullet \simeq 1.4 \times 10^6 M_\odot$ for NGC 4051
(Wandel, Peterson, \& Malkan 1999).
it is noted that their method tends to give
a smaller mass than the dynamical method (e.g., Koratkar \& Gaskell 1991).
For example, Hayashida et al. (1998) obtained $M_\bullet \simeq
9.16 \times 10^6 M_\odot$, $1.90 \times 10^7 M_\odot$, and
 $1.79 \times 10^7 M_\odot$ for NGC 4151, NGC 5548, and 3C 273,
respectively. On the other hand, Koratkar \& Gaskell (1991)
obtained $M_\bullet \simeq
3.2 \times 10^7 M_\odot$, $2.2 \times 10^8 M_\odot$, and
$3.5 \times 10^8 M_\odot$ for the same galaxies.
Although we do not know which method is more reliable, it will be
necessary to estimate the black hole mass for both BLS1s and NLS1s
with a consistent manner.

\section{STATISTICAL CONSIDERATION}

\subsection{Frequency of Occurrence of NLS1s}

Stephens (1989) show that NLS1s share $\approx$ 10\% of S1s
based on optical spectroscopy of 65 hard X-ray selected AGN.
The number of NLS1s has been increasing by means of soft X-ray
surveys of AGN (Puchnarewicz et al. 1992, 1996; BBF96; Xu et al. 1999).
Indeed, NLS1s comprise about half of the AGN in soft X-ray selected
samples (Grupe et al. 1999; Edelson et al. 1999). However,
since soft X-ray photons are easily absorbed by dense 
gas clouds in nuclear 
regions of Seyfert galaxies (e.g., Awaki et al. 1991),
any soft X-ray selected samples tend to miss such obscured AGN, 
giving rise to the overabundance of both NLS1s and BLS1s.
Hard X-ray, radio, or optical surveys are more useful 
for any statistical considerations.
Therefore, we adopt the fraction obtained by Stephens (1989).

\subsection{Interpretation Based on the Pole-on View Model}

If we adopt the pole-on view model for NLS1s, the fraction of NLS1s gives 
a critical viewing angle to NLS1s statistically by the following relation,

\begin{eqnarray}
{N_{\rm NLS1} \over {N_{\rm S1} + N_{\rm S2}}}
 & = & {N_{\rm NLS1} \over {N_{\rm NLS1} + N_{\rm BLS1} + N_{\rm S2}}}
 \nonumber \\
 & = & 1 - {\rm cos} ~ i_{\rm cr, NLS1}
\end{eqnarray}
where 
$N_{\rm NLS1}$, $N_{\rm BLS1}$, and $N_{\rm S2}$ 
are the numbers of NLS1s, BLS1s, and S2s,
respectively.
According to Stephens (1989), we have 

\begin{equation}
{N_{\rm NLS1} \over N_{\rm S1}} = 
{N_{\rm NLS1} \over {N_{\rm NLS1} + N_{\rm BLS1}}} = 0.1.
\end{equation}
The number ratio between S1s and S2s is obtained by the following surveys;
0.125 (Osterbrock \& Shaw 1988), 0.20 (Salzer 1989), 
and 0.435 (Huchra \& Burg 1992).
Note that these ratios are corrected for the completeness of 
the individual surveys.
The simple average of the three surveys gives $0.25 \pm 0.13$.
It is noted that this ratio gives a critical viewing angle to BLS1s is
$i_{\rm cr, BLS1} \simeq 37^\circ$. This value is almost consistent with the
opening angle of NLRs; $26^\circ \pm 11^\circ$ (Pogge 1989),
$32^\circ \pm 8^\circ$ (Wilson \& Tsvetanov 1994),  and
$29^\circ \pm 9^\circ$ (Schmitt \& Kinney 1996).
Therefore, the above average ratio appears reasonable.
Since NLS1s are included in the samples of S1s, we adopt 

\begin{equation}
{N_{\rm S1} \over N_{\rm S2}} = 
{N_{\rm NLS1} + N_{\rm BLS1} \over N_{\rm S2}} = 0.25.
\end{equation}

Using equations (1), (2), and (3), we obtain
$i_{\rm cr, NLS1} \simeq 10.9^\circ$.
Although this value is slightly larger than the previous estimate
based on the kinematical considerations [$4.8^\circ$ for FWHM(BLR) = 
1000 km s$^{-1}$ and $9.6^\circ$ for FWHM(BLR) = 2000 km s$^{-1}$],
there may be no serious inconsistency between the two estimates if
the observational ambiguities in both the estimates are taken 
into account.

\subsection{Interpretation Based on the Low-M$_\bullet$ Model}

If we adopt the low-$M_\bullet$ model, the number ratio between NLS1s and 
BLS1s can be attributed to the mass function of SMBHs in Seyfert nuclei.
Therefore, a direct way to prove this model is to show that NLS1s
have systematically lower black hole masses than BLS1s.
However, as mentioned in section 5.3, 
there is no reliable comparison of the black hole mass between
NLS1s and BLS1s. 
Recent high-resolution optical spectroscopy observations have shown that 
SMBHs with $M_\bullet \sim 10^{6-9} M_\odot$ are present in nearby galaxies
and there is a relationship between the SMBH mass and the spheroidal mass
of galaxies; i.e., $M_\bullet \approx 0.006 M_{\rm sph}$ 
(Kormendy et al. 1998).
It will become important to compare the spheroidal mass between NLS1s' 
and BLS1s' host galaxies.

\section{A NEW MODEL}

\subsection{Introduction}

As described in section 4,
it is hard to discriminate which model is more plausible among the two,
the pole-on view model or the low-$M_\bullet$ one.
Here it is remembered that there are two important correlations (section 2);
[O7] the correlation between Fe {\sc ii}/H$\beta$
intensity ratio and the FWHM of H$\beta$ emission; 
weak Fe {\sc ii} emitters always have larger FWHMs but stronger 
Fe {\sc ii} emitters have either larger or smaller FWHMs, and 
[X2] the correlation between the soft X-ray spectral index
and the FWHM of H$\beta$ emission;
S1s with larger FWHM(H$\beta$) (i.e., BLS1s) always have flat 
spectra but S1s with smaller FWHM(H$\beta$) (i.e., NLS1s)
emitters have either steep or flat spectra.

If we adopt the low-$M_\bullet$,
a tunable parameter is only the black hole mass. 
It seems difficult to explain the above two correlations
by tuning only $M_\bullet$. On the other hand, 
the pole-on view model has the viewing-angle dependence which is 
known to affect the visibility of some emission line components
because of the effect of selective obscuration.
Although it is natural that there is a scatter in  $M_\bullet$
from galaxy to galaxy, we try to find a possible model adopting 
the viewing angle as the primary parameter.

\subsection{The Fe II Emitting Region}

We begin our model with a question; ^^ ^^ Where is the Fe {\sc ii}
emitting region ?" The reason for this is that the Fe {\sc ii}
emission is unusually strong in NLS1s with respect to BLS1s [O7].
It has been often considered that 
the Fe {\sc ii} emission arises from partly ionized regions
heated by X-ray photons (Ferland \& Netzer 1979; Kwan \& Krolik 1981).
Since S2s show no Fe {\sc ii} features in the optical spectra,
the current unified model for Seyfert nuclei suggests that the
Fe {\sc ii} emitting regions should be inside the dusty tori.
Therefore, the most natural place is considered to be optically-thick
ionization-bounded clouds located in the BLR (Collin-Souffrin \&
Lasota 1988; Collin-Souffrin, Hameury, \& Joly 1988;
Marziani et al. 1996; Dultzin-Hacyan et al. 1999).

One of the important observational properties of NLS1s is the
evidence for the blue wing emission in UV emission lines, Ly$\alpha$
and C {\sc iv}$\lambda$1549 [U2].
Since their line widths amount to $\approx$ 5000 km s$^{-1}$,
there is a highly-ionized BLR (hereafter HI-BLR) even in NLS1s.
It has been suggested that the HI-BLR is associated with the jet-like
BLR (Marziani et al. 1996; Dultzin-Hacyan et al. 1999; see also
Sulentic et al. 1995). 
Marziani et al. (1996) present the first direct comparison between the 
HI-BLR (e.g., C~{\sc iv})
and the low-ionization BLR (hereafter LI-BLR) (e.g., H$\beta$)
for a sample of 52 radio-quiet AGN.
They found the following correlations;
the C~{\sc iv}
equivalent width EW(C~{\sc iv}) decreases with (a) increasing blueshift
of C~{\sc iv}  relative to H$\beta$ (broad component), (b) 
increasing  strength of the optical Fe~{\sc ii} multiplets, and 
(c) increasing strength of the optical-UV continuum.
These correlations and the systematic nature of the C~{\sc iv} blueshift
relative to H$\beta$  suggest a model where: (i) C~{\sc iv}
emitting clouds show a predominance of radial motion  (outflow because
it is assumed that the far side is obscured) in a bi-conical (and/or disk
wind) structure with a wide opening angle; i.e., the jet-like BLR. 
(ii) Fe~{\sc ii} optical emission
arises in a flattened distribution (possibly an accretion disk);
i.e., the disk-like BLR, and
(iii) H$\beta$ arises in a different and less flattened distribution.

Taking these properties into account, we adopt the following simple
two-component BLR model in the later discussion;
1) the disk-like BLR in which Fe {\sc ii} and H$\beta$ arise, and
2) the jet-like BLR in which C {\sc iv} and Ly$\alpha$ arise
(see Figure 1).
Even in the disk BLR, the gas is distributed in a clumpy form
because the volume filling factor in the BLR is generally low;
e.g., $\sim 10^{-7}$ (Peterson 1997).
If we assume that each BLR cloud has a spherical form, the outer surface
facing to the central engine is highly ionized (e.g., Pier \& Voit 1995).
Therefore, it is considered that the Fe {\sc ii} emitting region
is partly ionized zones which are shaded by highly- and 
intermediately-ionized zones from the central engine.
A possible formation mechanism of  partly-ionized regions
in the disk-like BLR will be discussed in section 7.4.

\begin{figure*}
\epsscale{1.0}
\plotone{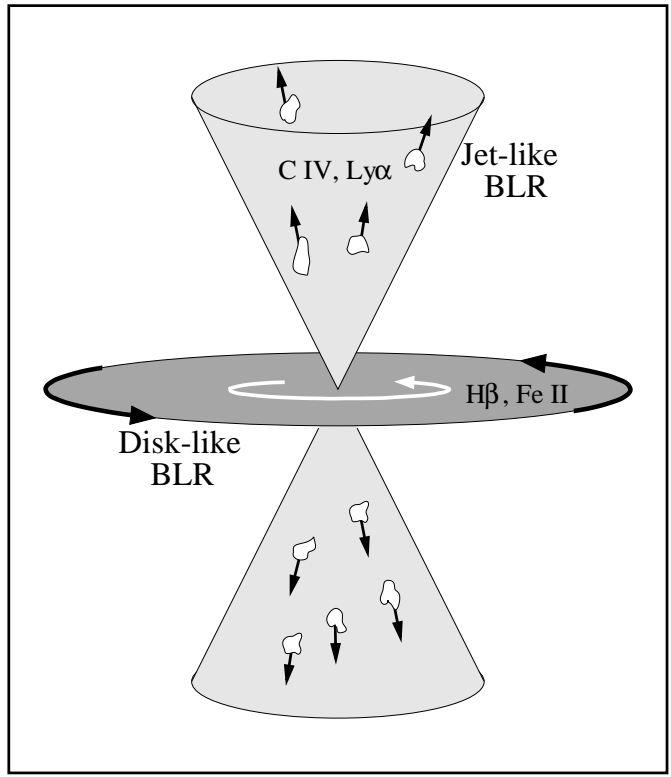}
\caption{%
A schematic illustration of the two-component BLR model.
This is reproduced from Figure 3 in Dultzin-Hacyan et al. (1999).
\label{fig1}
}
\end{figure*}

There is no direct information about the optical depth
from the central engine toward the Fe {\sc ii} emitting region.
However, we can infer it using observations and theoretical 
considerations on infrared Ca {\sc ii} triplet lines at 
8498 \AA, 8542 \AA, and 8662 \AA ~ (Persson 1988; Ferland \&
Persson 1989; Joly 1989) which are also considered to arise from 
the same partly-ionized regions as those of the Fe {\sc ii} emission
(Dultzin-Hacyan et al. 1999; Taniguchi et al. 1999).
Comparing photoionization models with the observations,
Ferland \& Persson (1989) found that optically thin clouds
with column densities $N_{\rm H} \sim 10^{23}$ cm$^{-2}$ cannot
reproduce the observed Ca {\sc ii} triplet line ratios
regardless of density and ionization parameters.
The observed Ca {\sc ii} spectra can be reproduced if
$N_{\rm H} \gtrsim 10^{24.5}$ cm$^{-2}$.
Therefore, the visibility of the partly-ionized regions
is expected to be strongly viewing-angle dependent.
If we observe this disk-like BLR from an inclined viewing angle
(e.g., $i_{\rm view} = 30^\circ$), the Fe {\sc ii} emitting region
located in the far-side half disk cannot be seen entirely
because of the large optical depth. On the other hand, the H$\beta$
emission can be seen from the entire BLR disk
because the ionized hydrogen is located in 
the outer surfaces of BLR clouds. On the other hand,
if we observe the disk-like BLR from a nearly pole-on view,
we can see both Fe {\sc ii} and H$\beta$, resulting in a higher
Fe {\sc ii}/H$\beta$ ratios together with narrower line width
with respect to those observed from an inclined viewing angle
(see Figure 1).
This explains the anti-correlation between the Fe {\sc ii}/H$\beta$
ratio and the FWHM(H$\beta$) [O7]. However, we note that 
the present model explains only a factor of two difference
in the Fe {\sc ii}/H$\beta$ ratio while the observed ratio
lies in a wider range between $\sim$ 0.1 -- $\sim$ 3
(e.g., Joly 1991). As mentioned before, the Fe {\sc ii}
emitting regions are thought to be very optically thick.
In order to explain the observed range, sophisticated
photoionization models will be necessary (see section 7.4).

Another concern is that the present model
cannot explain directly why there are NLS1s with both weaker
Fe {\sc ii}/H$\beta$ ratios and narrower line widths
although it explains the anti-correlation [O7].
Here it is again noted that the Fe {\sc ii} emitting regions 
are expected to be optically very thick.
It seems difficult to observe
the Fe {\sc ii} emitting regions in some NLS1s
 even if we see the disk-like BLR
from a nearly pole-on view. Alternatively, 
some NLS1s may have a smaller volume of the partly-ionized
region in the BLR. Although we cannot rule out 
another possibility that a variety in FWHM(H$\beta$) 
is partly controlled by the mass of SMBHs; i.e., 
some S1s have lower $M_\bullet$ than typical S1s
(the low-$M_\bullet$ model), it is not necessarily to introduce 
this idea to explain the observed relationship between the 
Fe {\sc ii}/H$\beta$ ratio and FWHM(H$\beta$).

\subsection{The Soft X-ray Emitting Region}

In this section we consider where the major emitting region 
of soft X-ray photons is. Several mechanisms may be 
responsible for the excess production of soft X-ray emission.
It has been often considered that the soft X-ray emission arises from 
the accretion disk (Madau 1988; Ross, Fabian, \& Mineshige 1992;
Ross \& Fabian 1993). For example, in an optically-thick,
geometrically-thin accretion disk, Madau (1988) take the occultation of
the innermost disk region due to self-shadowing and the reflection
effect of photons off the funnel wall into account and then find that 
UV and soft X-ray emission is enhanced if we observe this accretion disk
from a nearly pole-on view. However, Walter \& Fink (1993)
suggested that spectral energy distributions between UV and soft X ray
predicted by such accretion disk models appear inconsistent with 
the observations.

One of very important observational properties of NLS1s
related to the soft X-ray emitting region is the discovery
of evidence for the relativistic outflow in three NLS1s;
1H 0707$-$495, IRAS 13224$-$3809, and PG 1404+226
(Leighly et al. 1997). 
These outflows are probed by
the broad absorption features around 1 keV; 
the inferred blueshifts lie in a range
between 0.2$c$ and 0.57$c$.
In addition to these three,
Vaughan et al. (1999) find other three NLS1s (Ton S180, PG 1244+026,
and Ark 564) also have such
unusual absorption features. Moreover, there are three NLS1s 
with usual warm absorber (NGC 4051, IRAS 17020+45, and 
IRAS 20181$-$22); O {\sc vii} and
O {\sc viii} edges at 0.74 and 0.87 keV.
In summary, among the 22 NLS1s analyzed by Vaughan et al. (1999),
the above nine NLS1s show absorption features.
See Iwasawa, Brandt, \& Fabian (1998) for the X-ray absorption
in one of the strong-Fe {\sc ii} NLS1s, Mrk 507.
It is remarkable that the NLS1s with the absorption
have narrower H$\beta$ line widths;
an average FWHM $\approx$ 800 $\pm$ 180 km s$^{-1}$. On the other hand,
the remaining NLS1s have
an average FWHM $\approx$ 1290 $\pm$ 320 km s$^{-1}$.
We examine the difference of the frequency distribution of FWHMs
between the NLS1s with absorption features and NLS1s without them.
We adopt the null hypothesis that the NLS1s with and without absorption
features come from the same underlying distribution.
Applying the Kormogrov-Smirnov (KS) statistical test, we obtain
the probability of randomly selecting the FWHMs from the same
underlying population as $5.4 \times 10^{-5}$. Therefore, the difference
in FWHM between the two samples appears statistically real.
Vaughan et al. (1999) suggested that this property can be understood
in terms of the pole-on view model; e.g., if the absorbing material 
originates in an outflow from the disk, it would only be seen in 
low inclination systems.

Since relativistic outflows of plasma are expected to produce
synchrotron emission in the radio, they may be identical to 
relativistic radio jets. Indeed,
nuclear radio jets have been found in most Seyfert galaxies
(e.g., Ulvestad \& Wilson 1989).
Furthermore, recent detailed morphological studies of inner regions 
of the NLR in some nearby Seyfert nuclei
have shown that the optical NLRs are associated with the radio
jet (Bower et al. 1995; Capetti et al. 1995, 1996). These
observations have strongly suggested that the NLR associated with
the radio jet may be formed by the ionizing fast shock driven by the
radio jet rather than the photoionization (Dopita \& Sutherland
1995, 1996; Dopita et al. 1997; Bicknell et al. 1998; Falcke,
Wilson, \& Simpson 1998;
Wilson \& Raymond 1999; see also Daltabuit \& Cox 1972;
Wilson \& Ulvestad 1983; Norman \& Miley 1984).
Although it is still uncertain that the majority
of the NLR of Seyfert nuclei are formed by the ionizing shock 
(Laor 1998; Evans et al. 1999),
the spatial coincidence between the radio jets and the optical
emission-line gas means that the ionizing shock works in part.
If the radio jet interacts with the dense ambient gas
in very close to the central engine,
this can be responsible for the formation of hot plasma,
giving rise to the production of X-ray photons; note that
the energy of 0.1 keV corresponds to the kinetic temperature
$T_{\rm kin} \simeq 1.16 \times 10^6$ K. 

Here we consider what happens when a radio jet interacts with
the dense ambient gas following Norman \& Miley (1984). The jet is
characterized by the jet luminosity $L_{\rm jet}$, the jet
velocity $v_{\rm jet}$, and the solid opening angle of the jet
$\Omega_{\rm jet}$. The pressure exerted on the ambient gas
by the radio jet is estimated as

\begin{eqnarray}
p_{\rm jet} & \sim & 0.01
\left( {L_{\rm jet} \over {10^{44} ~ {\rm erg ~ s^{-1}}}} \right)
\left( {\Omega_{\rm jet}/4\pi \over {0.01}} \right)^{-1} \nonumber \\
 &  & \times\left( {r_{\rm jet} \over {1 ~ {\rm pc}}} \right)^{-2}
\left( {v_{\rm jet} \over {10^5 ~ {\rm km ~ s^{-1}}}} \right)^{-1}
~ {\rm dyne ~ cm}^{-2}.
\end{eqnarray}

\noindent where $r_{\rm jet}$ is the radial distance of the jet.
If we assume that the ambient gas clouds can cool and reach
pressure equilibrium in the cocoon of the jet,
we obtain a kinetic temperature of the gas

\begin{eqnarray}
T_{\rm kin} & \sim  & 10^4
\left( {n_{\rm e} \over {10^{10} ~ {\rm cm^{-3}}}} \right)^{-1}
\left( {L_{\rm jet} \over {10^{44} ~ {\rm erg ~ s^{-1}}}} \right)
\left( {\Omega_{\rm jet}/4\pi \over {0.01}} \right)^{-1}
\nonumber \\
 & & \times
\left( {r_{\rm jet} \over {1 ~ {\rm pc}}} \right)^{-2}
\left( {v_{\rm jet} \over {10^5 ~ {\rm km ~ s^{-1}}}} \right)^{-1}
~ {\rm K}
\end{eqnarray}

\noindent where $n_{\rm e}$ is the electron density.
For typical Seyfert nuclei, $L_{\rm jet}$ is of the order of
$10^{40}$ erg s$^{-1}$ at most (e.g., Wilson, Ward, \& Haniff 1988).
Although large-scale radio jets (i.e., kpc jets) have 
jet velocities of  the order of 10$^4$ km s$^{-1}$
(e.g., Wilson \& Ulvestad 1983;
Gallimore, Baum, \& O'dea 1996), the inner jet velocities
inferred from the broad absorption features in soft X-ray spectra
is of the order of 10$^5$ km s$^{-1}$ (Leighly et al. 1997).
An important question is where the soft X-ray emitting region is
along the jet. The soft X-ray excess is generally observed in
S1s (e.g., Mushotzky, Done, \& Pounds 1993). The unified model
suggests that the soft X-ray emitting region is hidden by
a dusty torus in S2s. Since the typical half height of dusty
tori is of the order of 0.1 pc (Taniguchi \& Murayama 1998 and
references therein), we estimate a typical radial distance of the
soft X-ray emitting region is $\sim$ 0.01 pc or less;
we thus adopt $r_{\rm jet} \sim r_{\rm soft~X} \sim$ 0.01 pc.
This is almost comparable to that of the disk BLR.
However, it is noted that 
the jet develops toward a direction perpendicular to
the accretion disk (i.e., to the BLR disk). Therefore,
electron densities in such ambient matter 
may be lower than those in the disk BLR and be similar 
to those in the warm absorber region (WAR);
we thus adopt the typical electron density of the WAR,
$\sim 10^8$ cm$^{-3}$ (e.g., Nishiura \& Taniguchi 1998 and 
references therein).
Then we obtain a typical kinetic temperature of the jet-driven
shocked region

\begin{eqnarray}
T_{\rm kin} & \sim & 10^6
\left( {n_{\rm e} \over {10^{8} ~ {\rm cm^{-3}}}} \right)^{-1}
\left( {L_{\rm jet} \over {10^{40} ~ {\rm erg ~ s^{-1}}}} \right)
\left( {\Omega_{\rm jet}/4\pi \over {0.01}} \right)^{-1}
\nonumber \\
 & & \times
\left( {r_{\rm soft~X} \over {0.01 ~ {\rm pc}}} \right)^{-2}
\left( {v_{\rm jet} \over {10^5 ~ {\rm km ~ s^{-1}}}} \right)^{-1}
~ {\rm K}.
\end{eqnarray}

\noindent This temperature is high enough to produce soft X-ray photons
and appears consistent with the estimated black-body temperatures for 
the NLS1s, $\sim$ 0.1 keV (Pounds et al. 1995; Puchnarewicz et al. 1995a,
1995b; Vaughan et al. 1999).
We note that we can obtain $T_{\rm kin} \sim 10^6$ K if we adopt 
$r_{\rm jet} \sim$ 0.001 pc and $n_{\rm e} \sim 10^{10}$ cm$^{-3}$.

We investigate whether or not this idea is responsible for the 
steep soft X-ray spectra observed in NLS1s.
Our model assumes that the excess soft X-ray emission is attributed to
the thermal emission made by jet-driven shocks.
Therefore, the soft X-ray spectrum is characterized by the black body
radiation with $T_{\rm kin} \sim 10^6$ K.  
Since the soft X-ray spectra of NLS1s are dominated by the so-called
soft excess emission over the underlying power-law continuum
(e.g., Vaughan et al. 1999), it seems reasonable to assume that 
the specific flux in the soft X-ray is described by the Planck function;

\begin{equation}
f_\nu(T) \approx B_\nu(T) = {2h \over c^2}
{{\nu^3} \over {e^{h\nu/kT} -1}}
\end{equation}
where $T \equiv T_{\rm kin}$.
The spectral index at a frequency $\nu$ is given by

\begin{equation}
\alpha(\nu,T) = {dB_\nu(T)/d\nu \over B_\nu(T)/\nu}
= 3 - {{h\nu/kT} \over {1 - e^{-h\nu/kT}}}.
\end{equation}
Since $f_\nu/(h\nu) \propto \nu^{\alpha - 1} \propto E^{-\Gamma}$,
the photon index $\Gamma$ is related to the spectral index as

\begin{equation}
\Gamma(\nu,T) = 1 - \alpha(\nu,T)
= {{h\nu/kT} \over {1 - e^{-h\nu/kT}}} - 2.
\end{equation}
Introducing the following parameter 

\begin{equation}
x \equiv {h\nu \over kT} = 11.6 \left({E \over {\rm keV}} \right)
\left({T \over 10^6 {\rm K}} \right)^{-1},
\end{equation}
we obtain a relation

\begin{equation}
{T \over 10^6 {\rm K}} = {11.6 \over x} \left({E \over {\rm keV}} \right).
\end{equation}
In Table 2, we give values of $\Gamma$ as a function of $T$
at an energy of $E$ = 1 keV; i.e., $\Gamma \simeq \Gamma_{\rm soft}$. 
It is known that
the NLS1s have $\Gamma_{\rm soft} \simeq$ 3 -- 5 [X1] (e.g., BBF96).
These steeper soft X-ray spectra can be explained 
if $T_{\rm kin} \simeq (1.7$ -- $2.3) \times 10^6$ K.
The spectral fitting for 22 NLS1s gives an average kinetic
temperature $\overline{T}_{\rm kin} = (2.00 \pm 0.86) \times 10^6$ K
(Vaughan et al. 1999). Therefore our model appears consistent with
the observations.

\begin{deluxetable}{ccc}
\tablenum{2}
\tablewidth{24em}
\tablecaption{Soft X-ray photon indices as a function of $T$
at an energy of $E$ = 1 keV}
\tablehead{
 \colhead{$\Gamma$} & \colhead{$x$\tablenotemark{a}}
 & $T$ ($10^6$ K) \nl
}
\startdata
1 &  2.8214 & 4.11 \nl
2 &  3.9027 & 2.96 \nl
3 &  4.9651 & 2.34 \nl
4 &  5.9849 & 1.94 \nl
5 &  6.9936 & 1.67 \nl
\enddata
\tablenotetext{a}{$x \equiv h\nu/(kT)$.}
\end{deluxetable}

In the above analysis, we assumed that the soft X-ray emitter is at rest.
However, the observed absorption features at 1.1 -- 1.4 keV imply that
the soft X-ray absorbers are moving at velocities of $\sim$ 0.2$c$ -- 0.6$c$
[X4]. Taking this point into account, one can obtain

\begin{equation}
{T_{\rm rest} \over 10^6 {\rm K}} = {11.6 \over x} 
\left({E_{\rm obs} \over {\rm keV}} \right)
\delta^{-1}
\end{equation}
where $T_{\rm rest}$ is the kinetic temperature in the rest frame
[note that this is the same as $T$ in equation (13)],
$E_{\rm obs}$ is the energy in the observed frame, and $\delta$
is the kinetic Doppler factor defined by

\begin{equation}
\delta \equiv [\gamma (1 - \beta ~ {\rm cos} ~ i_{\rm view})]^{-1},
\end{equation}
where $\beta$ is the bulk velocity of the emitting region
in units of the light speed (i.e, $\beta = v/c$) and
$\gamma = (1 - \beta^2)^{-1/2}$ is the Lorenz factor
(e.g., Ghisellini et al. 1993). 
Given the observed range of $\beta$ between 0.2 and 0.6,
the kinetic temperature of the soft X-ray emitting region
can be lowered at most by a factor between 1.1 -- 1.6.
In Table 3, we give a summary of the photon index at an energy 
of 1 keV as functions of $T_{\rm kin}$, $\beta$, and $i_{\rm view}$.
It is shown that 
the observed steep photon indices are obtained within reasonable
parameter ranges; $T_{\rm kin} \simeq$ (1 -- 2) $\times 10^6$ K
and $\beta \simeq$ 0.2 -- 0.7.

\begin{deluxetable}{cccccccc}
\tablenum{3}
\tablecaption{The photon index at 1 keV as functions of
              $T_{\rm kin}$, $\beta$, and $i_{\rm view}$}
\tablehead{
  \colhead{$\beta$} &
  \colhead{$i_{\rm view}$} &
  \colhead{$\delta$} &
  \multicolumn{5}{c}{$\Gamma$(1~keV)} \nl
  & (\arcdeg)
}
\startdata
 & & & \multicolumn{5}{c}{$T_{\rm kin}$ ($10^{6}$ K)} \nl
 & & & 1.0 & 2.0 & 3.0 & 4.0 & 5.0 \nl
\cline{4-8}
0.0 & \nodata & 1.000 & 9.61 & \phs3.82  & \phs1.95  & \phs1.07  & \phs0.57  \nl
0.1 &  0 & 1.106 & 8.50 & \phs3.28  & \phs1.61  & \phs0.83  & \phs0.39  \nl
0.2 &  0 & 1.225 & 7.48 & \phs2.78  & \phs1.30  & \phs0.61  & \phs0.23  \nl
0.3 &  0 & 1.363 & 6.52 & \phs2.32  & \phs1.01  & \phs0.42  & \phs0.08  \nl
0.4 &  0 & 1.528 & 5.60 & \phs1.89  & \phs0.75  & \phs0.23  &   $-0.05$ \nl
0.5 &  0 & 1.732 & 4.71 & \phs1.47  & \phs0.50  & \phs0.06  &   $-0.18$ \nl
0.6 &  0 & 2.000 & 3.82 & \phs1.07  & \phs0.26  &   $-0.11$ &   $-0.31$ \nl
0.7 &  0 & 2.380 & 2.91 & \phs0.67  & \phs0.02  &   $-0.27$ &   $-0.43$ \nl
0.8 &  0 & 3.000 & 1.95 & \phs0.26  &   $-0.22$ &   $-0.44$ &   $-0.56$ \nl
0.9 &  0 & 4.359 & 0.86 &   $-0.19$ &   $-0.49$ &   $-0.63$ &   $-0.71$ \nl
\hline
0.1 &  5 & 1.105 & 8.50 & \phs3.28  & \phs1.61  & \phs0.83  & \phs0.39  \nl
0.2 &  5 & 1.224 & 7.48 & \phs2.78  & \phs1.30  & \phs0.62  & \phs0.23  \nl
0.3 &  5 & 1.361 & 6.53 & \phs2.33  & \phs1.02  & \phs0.42  & \phs0.08  \nl
0.4 &  5 & 1.524 & 5.62 & \phs1.89  & \phs0.76  & \phs0.24  &   $-0.05$ \nl
0.5 &  5 & 1.725 & 4.73 & \phs1.48  & \phs0.51  & \phs0.07  &   $-0.18$ \nl
0.6 &  5 & 1.989 & 3.85 & \phs1.08  & \phs0.27  &   $-0.10$ &   $-0.31$ \nl
0.7 &  5 & 2.360 & 2.95 & \phs0.69  & \phs0.03  &   $-0.26$ &   $-0.43$ \nl
0.8 &  5 & 2.955 & 2.01 & \phs0.28  &   $-0.21$ &   $-0.43$ &   $-0.56$ \nl
0.9 &  5 & 4.215 & 0.94 &   $-0.16$ &   $-0.47$ &   $-0.62$ &   $-0.70$ \nl
\hline
0.1 & 10 & 1.104 & 8.51 & \phs3.28  & \phs1.61  & \phs0.83  & \phs0.40  \nl
0.2 & 10 & 1.220 & 7.51 & \phs2.80  & \phs1.31  & \phs0.62  & \phs0.24  \nl
0.3 & 10 & 1.354 & 6.57 & \phs2.35  & \phs1.03  & \phs0.43  & \phs0.09  \nl
0.4 & 10 & 1.512 & 5.68 & \phs1.92  & \phs0.77  & \phs0.25  &   $-0.04$ \nl
0.5 & 10 & 1.706 & 4.81 & \phs1.52  & \phs0.53  & \phs0.08  &   $-0.17$ \nl
0.6 & 10 & 1.955 & 3.95 & \phs1.13  & \phs0.30  &   $-0.08$ &   $-0.29$ \nl
0.7 & 10 & 2.299 & 3.08 & \phs0.74  & \phs0.07  &   $-0.24$ &   $-0.41$ \nl
0.8 & 10 & 2.828 & 2.17 & \phs0.35  &   $-0.16$ &   $-0.40$ &   $-0.53$ \nl
0.9 & 10 & 3.835 & 1.18 &   $-0.06$ &   $-0.41$ &   $-0.57$ &   $-0.67$ \nl
\hline
0.1 & 20 & 1.098 & 8.57 & \phs3.31  & \phs1.63  & \phs0.84  & \phs0.40  \nl
0.2 & 20 & 1.207 & 7.62 & \phs2.85  & \phs1.34  & \phs0.64  & \phs0.25  \nl
0.3 & 20 & 1.328 & 6.74 & \phs2.42  & \phs1.08  & \phs0.46  & \phs0.12  \nl
0.4 & 20 & 1.468 & 5.91 & \phs2.03  & \phs0.84  & \phs0.29  &   $-0.01$ \nl
0.5 & 20 & 1.634 & 5.11 & \phs1.66  & \phs0.61  & \phs0.14  &   $-0.13$ \nl
0.6 & 20 & 1.834 & 4.34 & \phs1.30  & \phs0.40  &   $-0.01$ &   $-0.24$ \nl
0.7 & 20 & 2.087 & 3.58 & \phs0.96  & \phs0.20  &   $-0.15$ &   $-0.34$ \nl
0.8 & 20 & 2.417 & 2.84 & \phs0.64  & \phs0.01  &   $-0.28$ &   $-0.44$ \nl
0.9 & 20 & 2.825 & 2.18 & \phs0.36  &   $-0.16$ &   $-0.40$ &   $-0.53$ \nl
\hline
0.1 & 30 & 1.089 & 8.65 & \phs3.35  & \phs1.66  & \phs0.86  & \phs0.42  \nl
0.2 & 30 & 1.185 & 7.79 & \phs2.93  & \phs1.39  & \phs0.68  & \phs0.28  \nl
0.3 & 30 & 1.289 & 7.01 & \phs2.55  & \phs1.16  & \phs0.52  & \phs0.16  \nl
0.4 & 30 & 1.402 & 6.28 & \phs2.20  & \phs0.95  & \phs0.37  & \phs0.05  \nl
0.5 & 30 & 1.527 & 5.60 & \phs1.89  & \phs0.75  & \phs0.23  &   $-0.05$ \nl
0.6 & 30 & 1.665 & 4.97 & \phs1.59  & \phs0.58  & \phs0.11  &   $-0.15$ \nl
0.7 & 30 & 1.814 & 4.41 & \phs1.34  & \phs0.42  & \phs0.00  &   $-0.23$ \nl
0.8 & 30 & 1.953 & 3.96 & \phs1.13  & \phs0.30  &   $-0.08$ &   $-0.29$ \nl
0.9 & 30 & 1.976 & 3.89 & \phs1.10  & \phs0.28  &   $-0.09$ &   $-0.30$ \nl
\enddata
\end{deluxetable}

Another merit of our model is that 
the relativistic outflows also lead to the enhancement of 
the soft X-ray luminosity due to 
the kinematical Doppler effect; i.e.,

\begin{equation}
L_{\rm soft}^{\rm obs} = \delta^4 ~ L_{\rm soft}^{\rm rest}.
\end{equation}
Therefore NLS1s viewed from a more pole-on 
tend to have higher soft X-ray luminosities given a certain 
relativistic jet velocity. This explains why soft X-ray surveys
tend to pick up NLS1s more preferentially.
In Table 4, we give numerical values of $\delta^4$ for a 
various set of both $\beta$ and $i_{\rm view}$.
It is shown that the luminosity brightening due to the 
kinematic Doppler factor for the pole-on view is 
larger by a factor of $\approx$ 2.08 (1.15)
than that for $i_{\rm view} = 30^\circ$
if $\beta = 0.6$ ($\beta = 0.2$).
Therefore, our model suggests that NLS1s have steeper photon indices 
in the soft X-ray than BLS1s. It is also suggested that
NLS1s are systematically brighter in the soft X-ray than BLS1s.
All these are consistent with the observations ([X1] and [X2]).
NLS1s with both smaller $\Gamma$ and narrower FWHM(H$\beta$) may be 
either those with higher (or very lower) kinetic temperatures or 
those with higher-velocity outflows or both.
Furthermore, our model appears consistent with the fact that
the detection rate of typical WAR is lower in the NLS1s than
that in the BLS1s [X4].

\begin{deluxetable}{cccccc}
\tablenum{4}
\tablewidth{40em}
\tablecaption{%
The luminosity brightening factor ($\delta^4$)
due to the kinematic Doppler effect
as functions of $\beta$ and $i_{\rm view}$}
\tablehead{
   \colhead{$\beta$} & \multicolumn{5}{c}{$\delta^4$} \nl
 & \colhead{$i_{\rm view} = 0^\circ$} & \colhead{$i_{\rm view} = 5^\circ$} &
   \colhead{$i_{\rm view} = 10^\circ$} & \colhead{$i_{\rm view} = 20^\circ$} &
   \colhead{$i_{\rm view} = 30^\circ$} 
}
\startdata
0.1 & 1.494 & 1.491 & 1.484 & 1.454 & 1.408 \nl
0.2 & 2.250 & 2.241 & 2.216 & 2.119 & 1.972 \nl
0.3 & 3.449 & 3.427 & 3.361 & 3.114 & 2.759 \nl
0.4 & 5.444 & 5.390 & 5.229 & 4.650 & 3.867 \nl
0.5 & 9.000 & 8.864 & 8.473 & 7.121 & 5.443 \nl
0.6 & 16.00 & 15.64 & 14.62 & 11.32 & 7.691 \nl
0.7 & 32.11 & 31.00 & 27.93 & 18.96 & 10.82 \nl
0.8 & 81.00 & 76.25 & 63.97 & 34.13 & 14.56 \nl
0.9 & 361.0 & 315.5 & 216.2 & 63.72 & 15.25 \nl
\enddata
\end{deluxetable}

We investigate in more detail the important correlation [X2].
First, we consider why S1s with larger FWHM(H$\beta$) (i.e., BLS1s) 
have flatter spectra. 
Since  BLS1s are viewed from intermediate orientations,
the brightening due to the kinematic Doppler effect is generally weak.
Therefore the soft X-ray excess emission is less dominant than that
in NLS1s. We also have to take account that BLS1s with larger
FWHM(H$\beta$) tend to be observed from more inclined viewing angles
if the intrinsic FWHM(H$\beta$) is not so different from S1 to S1.
Thus, the extinction of soft X-ray photons due to dust grain above the dusty
tori are expected to more serious on average for the BLS1s than for the NLS1s.
These two orientation effects can be responsible for the flatter
soft X-ray spectra of the BLS1s.

Second, we consider why NLS1s with flat soft X-ray spectra.
If the jet velocity is not so highly relativistic although the jet
can cause shocked regions with $T_{\rm kin} \sim 10^6$ K, or if the
volume of the shocked region is fairly small,
the blackbody radiation to the power law continuum luminosity ratio
is small. Since the viewing angle dependence due to the kinetic
Doppler effect is also small, the soft X-ray spectrum is dominated 
by the power-law continuum and thus $\Gamma_{\rm soft} \sim 2$
regardless of FWHM(H$\beta$). 

Thirdly, we consider why there is a large scatter in
$\Gamma_{\rm soft}$ of the NLS1s.
This observed scatter can be attributed either to 
a variety in $T_{\rm BB}$ (see Table 3) or to
the relative predominance of the black body radiation
with respect to the power-law continuum or both.
In order to investigate which effect is more important,
in Figure 2, we show a diagram between 
the soft X-ray photon index $\Gamma_{\rm soft} \equiv
\Gamma$(0.1 -- 2.4 keV) and the black-body to power-law luminosity
ratio in the X ray (i.e., 0.6 -- 10 keV) for the sample of 
NLS1s studied by Vaughan et al. (1999) who performed the spectral
fitting of the ASCA X-ray spectra using the two components;
the black body radiation and the power-law continuum.
Here a Hubble constant
$H_0$ = 50 km s$^{-1}$ Mpc$^{-1}$ and
a deceleration parameter $q_0 = 0.5$ are adopted.
The black body temperatures estimated by them are shown by
filled circles with different colors.
Note that a black body temperature of $T_{\rm BB} = 1.7 \times 10^6$ K
is assumed in the fitting for both I Zw 1 and Mrk 507 
because of relatively poor S/N of the spectra (Vaughan et al. 1999).
We construct two-component models in which 
the black body radiation and the power-law continuum
are taken into account. Here we adopt the canonical 
hard X-ray photon index for NLS1s $\Gamma_{\rm hard} = 2.1$
(Vaughan et al. 1999) for the power-law continuum.
Our model results are also shown in Figure 2
by thick curves as a function of $T_{\rm BB}$.
Although the spectral fitting by Vaughan et al. (1999) 
leads to a variety of $T_{\rm BB}$ by a factor of three
[i.e., $T_{\rm BB} \simeq$ (1 -- 3) $\times 10^6$ K],
our model results suggest that the majority of the NLS1s
have $T_{\rm BB} \simeq$ (0.7 -- 1) $\times 10^6$ K.
Although these values are slightly lower than our previous estimates
based on Table 3 [i.e., $T_{\rm BB} \simeq$ (1 -- 2) $\times 10^6$ K],
this slight difference is due to that 
our previous estimates do not include the 
contribution of the power-law continuum.
It is likely that $L_{\rm BB} / L_{\rm PL}$ seems to be evaluated
more accurately than $T_{\rm BB}$ in the spectral fitting by
Vaughan et al. (1999) because $L_{\rm BB}$ is estimated by integrating 
the soft excess  emission while the estimate of $T_{\rm BB}$
is sensitive to the peak in the soft X-ray part of the spectra
which has often poorer S/N.
Therefore, the observed scatter in $\Gamma_{\rm soft}$
of the NLS1s is attributed mainly to the relative predominance
of the black body radiation.

\begin{figure*}
\epsscale{1.0}
\plotone{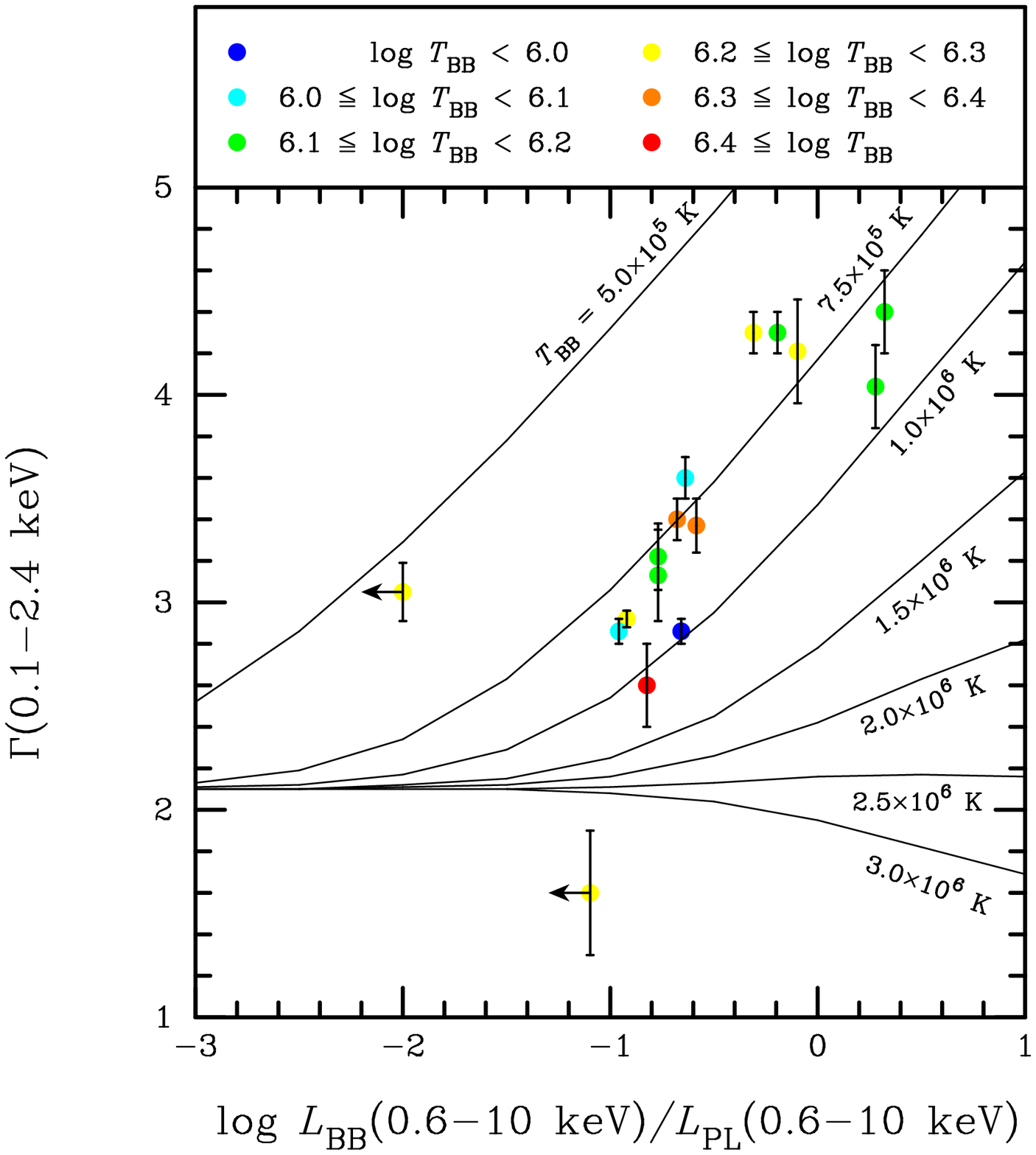}
\caption{%
A diagram between
the soft X-ray photon index $\Gamma_{\rm soft}$
and the black-body to power-law luminosity
ratio in the X ray (i.e., 0.6 -- 10 keV) for the sample of
NLS1s studied by Vaughan et al. (1999).
The black body temperatures estimated by them are shown by
filled circles  with different colors (see the upper panel).
Our model results by thick curves as a function of
$T_{\rm BB}$ in which the canonical hard X-ray photon index
for NLS1s is adopted to be 2.1 (Vaughan et al. 1999).
\label{fig2}
}
\end{figure*}

Here a question arises as what determines the black body
temperature. Since the origin of the black body radiation
is attributed to the jet-driven shocks in our model,
the temperature is determined by the ratio between the
mechanical luminosity of the jet and 
the gas mass of the shocked region if we neglect the 
relativistic effect of the jet velocity.
In Figure 3, we show a diagram between $\Gamma_{\rm soft}$
and the soft X-ray luminosity to the radio jet power
at $\lambda$ = 6 cm for the Vaughan et al.'s (1999) NLS1 sample. 
The radio continuum data are taken from Ulvestad et al. (1995).
The data set is too small to obtain any statistically
significant results. However, it is interesting to see a trend
that NLS1 with steeper $\Gamma_{\rm soft}$ have higher 
$L_{\rm soft}/P_{\rm jet}$ ratios [see also Figure 9 in Walter \&
Fink (1993)].  This tendency may be
interpreted by that NLS1s with a lot of dense ambient gas
have higher soft X-ray luminosities because of relatively larger
volume of the shocked region but the temperature of the shocked
region does not increase significantly, giving rise to 
steeper $\Gamma_{\rm soft}$. Unfortunately, the black body
temperature is available only for three NLS1s.
In order to verify this possibility,
more sensitive, soft and hard X-ray spectroscopy as well as deep 
radio-continuum mapping for a larger sample
of S1s will be recommended.
In Figure 4, we also show a diagram between $\Gamma_{\rm soft}$
and $L_{\rm soft}$ for the NLS1s studied by BBF96 (upper panel) and 
the S1s (NLS1s + BLS1s) studied by Walter \& Fink (1993)
(lower panel). It is shown that there is a clear, positive
correlation only for the NLS1s; its correlation 
coefficient is 0.66. 
This implies that NLS1s with higher $L_{\rm soft}$ have a larger
volume of the shocked region but have a lower black body temperature;
note that the soft X-ray luminosity due to the black body radiation
is estimated to be $L_{\rm soft}({\rm BB}) =4 \pi r^2 \sigma 
T_{\rm BB}^4$ where $\sigma$ is the Stephan-Boltzmann constant
if the shocked region has a spherical form with a radius of $r$.

\begin{figure*}
\epsscale{1.0}
\plotone{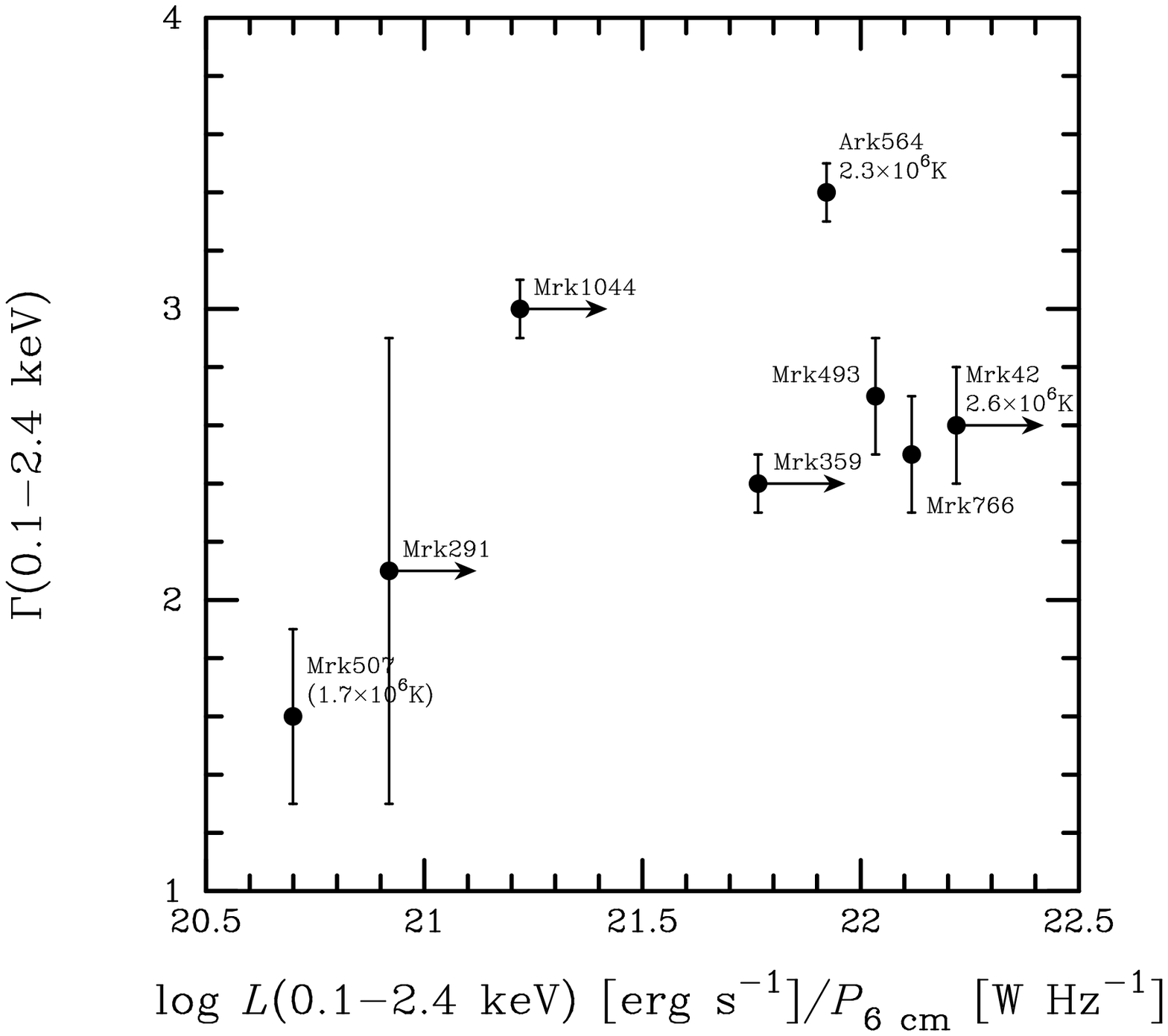}
\caption{%
A diagram between $\Gamma_{\rm soft}$
and the soft X-ray luminosity to the radio jet power
at $\lambda$ = 6 cm for the Vaughan et al.'s (1999) NLS1 sample.
\label{fig3}
}
\end{figure*}

\begin{figure*}
\epsscale{0.8}
\plotone{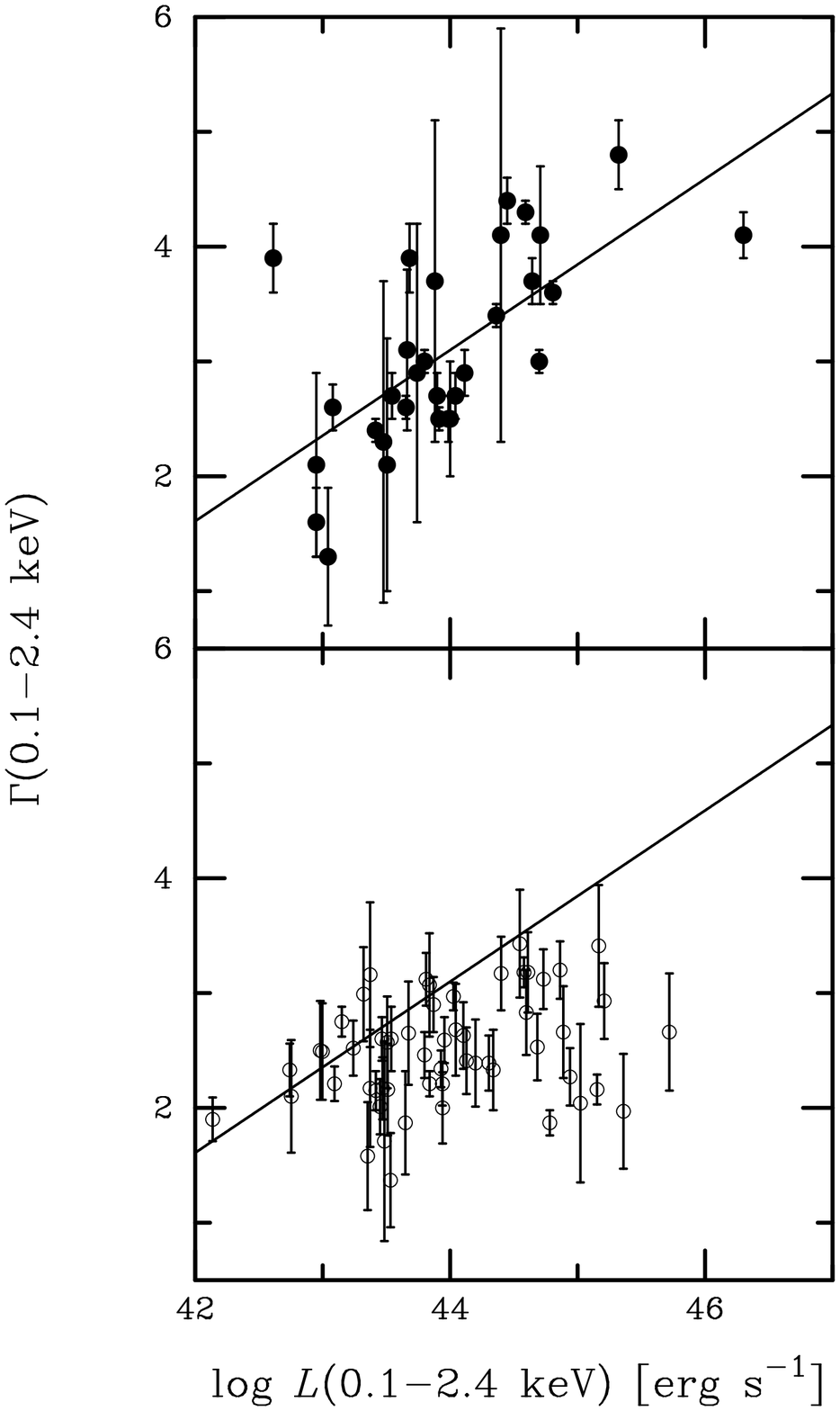}
\caption{%
A diagram between $\Gamma_{\rm soft}$
and $L_{\rm soft}$ for the NLS1s studied by BBF96 (upper panel)
and S1s studied by Walter \& Fink (1993) (Lower panel).
\label{fig4}
}
\end{figure*}

Finally we mention about a typical size of the soft X-ray emitting 
regions. The variability timescale observed in the soft X ray is  
of the order of one day or less (BBF96),
corresponding to a linear size of $\sim 3 \times 10^{15}$ cm
$\sim$ 0.001 pc. 
The soft X-ray emitting region is located at
a distance of $\sim$ 0.01 pc from the central engine in our model.
However, the above timescale suggests that 
a typical size of the soft X-ray emitting region
is as small as $\sim$ 0.001 pc.
As mentioned before, we have an alternative option;
the soft X-ray emitting region is located at
a distance of $\sim$ 0.001 pc from the central engine
and a typical size of the soft X-ray emitting region
is also as small as $\sim$ 0.001 pc.

\subsection{A New Photoionization Scenario for the 
Disk-like Broad Line Region}

We have described the pole-on view model is responsible 
for the two important correlations, [O7] (section 7.2) and 
[X2] (section 7.3). In this section, we investigate how these
two correlations are linked physically.

It is widely accepted that the BLR is photoionized by the nonthermal
continuum from the central engine although shock heating models 
are also sometimes discussed (e.g., Rees 1984; Blandford 1990).
As discussed in section 7.2, optically-thick, partly ionized regions
from which the Fe {\sc ii} emission arises are mainly heated
by X-ray photons (Ferland \& Netzer 1979; Kwan \& Krolik 1981).
High energy photons formed by the synchrotron self Compton effect
are associated with relativistic outflows and thus are thought 
to be highly beamed. The direction of the beaming is believed 
to be perpendicular to the accretion disk. Therefore, 
it is unlikely that these high energy photons are 
responsible for the ionization of the disk-like BLR. 
Indeed Wang et al. (1996) showed that NLS1s with strong Fe {\sc ii}
emission have lower X-ray luminosities at 2 keV and thus suggested
that the Fe {\sc ii} emitting region cannot be heated by X rays
with energies of $\sim$ 2 keV.
Instead, the optically-thick accretion disk around a supermassive 
compact object is also thought to play an important role in the 
ionization of the disk-like BLR.
The optically-thick condition leads to that gas particles
dissipate their energy locally because of the viscosity. In this case,
we can approximate the local emission as blackbody. Since the 
accretion disk has a temperature gradient, superposition of such
blackbody emission with various temperatures
can mimic to the power-law like continuum emission.
The temperature of such an accretion disk around a Schwarzschild 
supermassive black hole can be written as

\begin{eqnarray}
T(r) & \approx & 2.8 \times 10^5 (\dot{M} / \dot{M}_{\rm Edd})
M_{\bullet, 8}^{-1/4} [r / (3 r_{\rm S})]^{-3/4} ~~ {\rm K} ~~
\nonumber \\
 & & \makebox[11em]{}
({\rm for} ~~ r \geq 3 r_{\rm S}),
\end{eqnarray}

where $\dot{M}$ is the accretion rate, $\dot{M}_{\rm Edd}$ 
is the Eddington accretion rate, $M_{\bullet, 8}$ is the black hole 
mass in units of $10^8 M_\odot$, and $r_{\rm S}$ is the Schwarzschild
radius (e.g., Peterson 1997). Therefore, this radiation is mainly
emitted in the UV region, responsible for the photoionization of
the disk-like BLR. However, since this continuum emission
does not contain a lot of higher energy photons, it is unlikely that
this continuum creates a huge volume of partly ionized regions 
in the disk-like BLR (see upper panel of Figure 6).
Therefore, it is expected that a dense gas cloudlet immersed in the 
disk-like BLR consists of both a fully-ionized zone facing to the
continuum and a neutral zone. This means that Fe {\sc ii} emission 
cannot arise from this cloudlet efficiently. 

As shown in Figure 4, there is the clear correlation
between $\Gamma_{\rm soft}$ and the soft X-ray luminosity.
In addition, Wang et al. (1996) showed that there is a positive 
correlation between $\Gamma_{\rm soft}$ and
the Fe {\sc ii} $\lambda$4570/H$\beta$ ratio.
The simplest interpretation for these correlations is that
the stronger Fe {\sc ii} emission is attributed to the soft X-ray
excess emission; i.e., Fe {\sc ii} emitting regions in the
disk-like BLR are photoionized by the soft X-ray excess emission.
We have shown that jet-driven shocks lead to the formation of hot 
regions with $T_{\rm kin} \simeq 10^6$ K. Since 
this emission is approximated by the blackbody radiation,
soft X-ray photons are emitted isotropically; note, however, that
the shocked region are thought to be moving a relativistic velocity
and thus the radiation field is modulated by this effect.
These soft X-ray photons irradiate the disk-like BLR.
The outer surfaces of cloudlets in the disk-like BLR are 
photoionized by these photons. If these cloudlets are ionization-bounded,
there are neutral regions inside the cloudlets. 
Since some high energy photons can penetrate into the inner regions,
it is expected that partly ionized regions are made there.
In Figure 5, we show a diagram between the Fe {\sc ii}/H$\beta$ ratio
and the black-body radiation to power-law continuum luminosity ratio
in the hard X ray which is estimated by Vaughan et al. (1999).
Although the data points show a large scatter, there is a 
marginal tendency that NLS1s with larger Fe {\sc ii}/H$\beta$ ratios
tend to have higher $L_{\rm BB}/L_{\rm PL}$ ratios.
We thus propose that this is the formation mechanism of 
the Fe {\sc ii} emitting region in the disk-like BLR
(see lower panel of Figure 6).
Although we are not going to give more detailed photoionization
modeling, the new idea presented here appears to explain the 
origin of the Fe {\sc ii} emitting region qualitatively.

\begin{figure*}
\epsscale{1.0}
\plotone{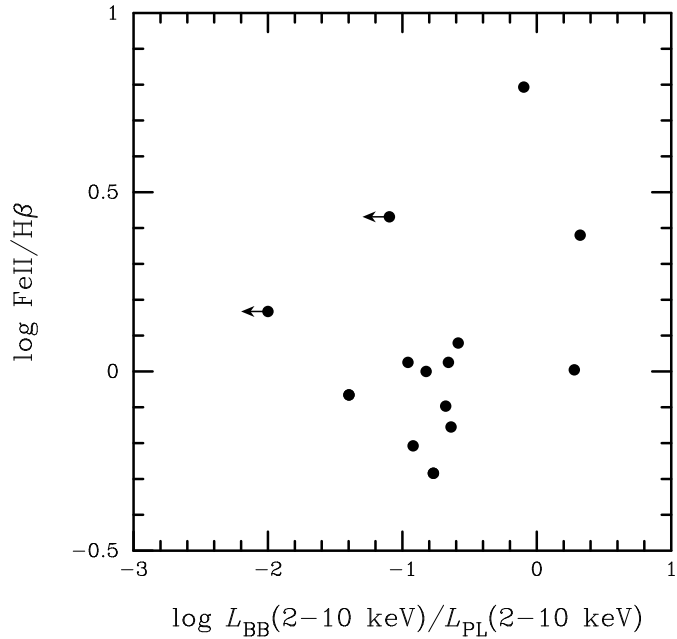}
\caption{%
A diagram between the Fe {\sc ii}/H$\beta$ ratio
and the black-body radiation to power-law continuum luminosity ratio
in the hard X ray which is estimated by Vaughan et al. (1999).
\label{fig5}
}
\end{figure*}

\begin{figure*}
\epsscale{1.0}
\plotone{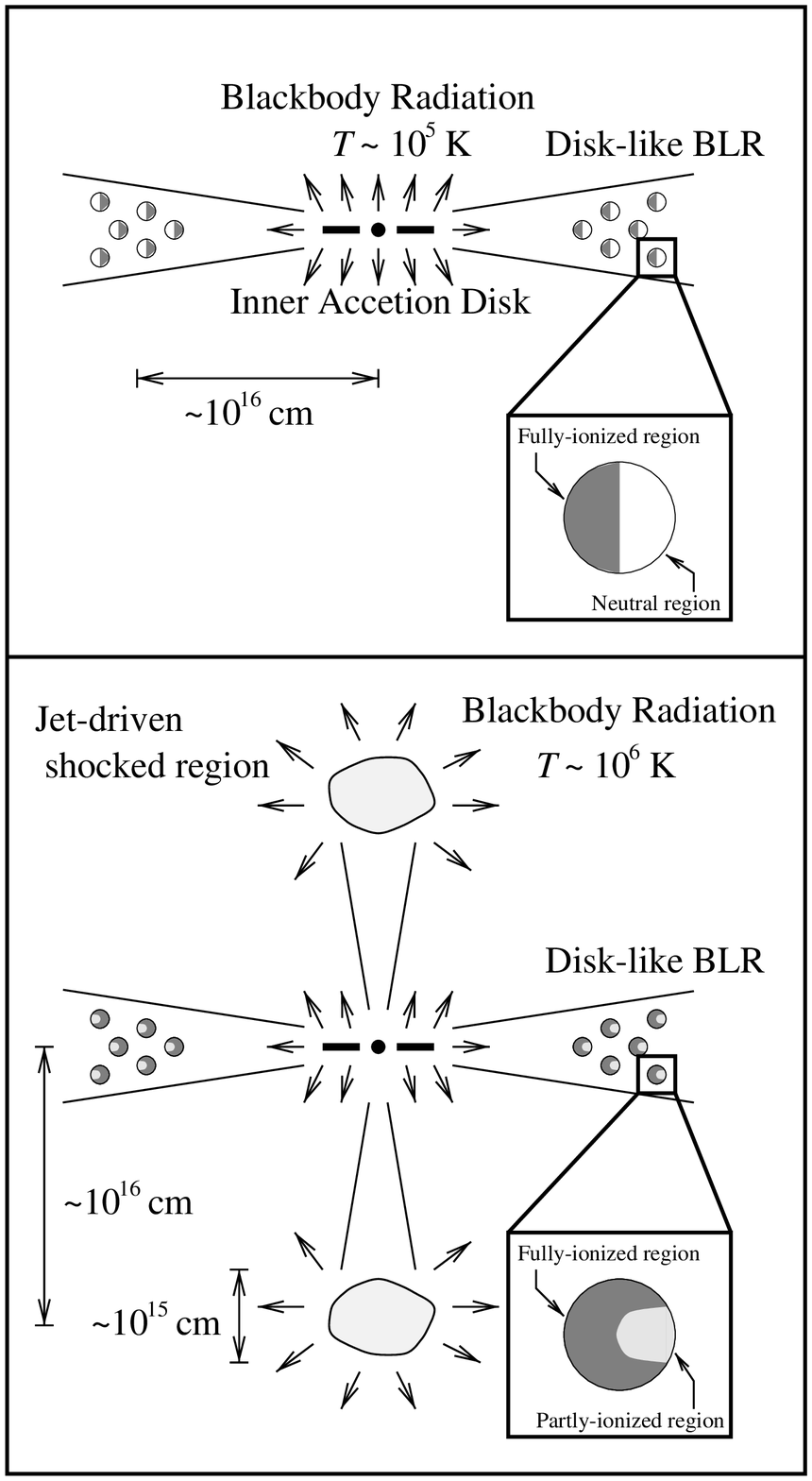}
\caption{%
Upper panel: Photoionization of cloudlets in the disk-like BLR by
the UV continuum emission from the inner accretion disk.
Lower panel: Photoionization of cloudlets in the disk-like BLR by
the additional soft X-ray sources, giving rise to the formation 
of partly ionized regions.
\label{fig6}
}
\end{figure*}

\subsection{Confrontation with Observation}

Our new model was constructed to explain the two most important 
correlations; between the Fe {\sc ii}/H$\beta$ intensity ratio
and the FWHM(H$\beta$) [O7] and  between the soft X-ray photon
index $\Gamma$ and the FWHM(H$\beta$) [X2].
Although these two correlations are explained by the viewing angle
dependence without invoking other mechanisms,
there are a number of other important observational properties of NLS1s 
summarized in section 2. Therefore, in this section, we investigate
whether or not our new model is also consistent with them.

The equivalent widths of Blamer lines of NLS1s are on average lower
by a factor of $\approx$ 2 than those of BLS1s [O3]. This is consistently
explained by the difference of the average viewing angle between
the NLS1s and the BLS1s if the optical continuum emission arises
mainly from the optically-thick accretion disk.
The optical polarization of NLS1s is interpreted by the dust scattering [O9].
This may be caused either at the inner wall of dusty tori or the 
dusty NLR clouds (Netzer \& Laor 1993; Wolf \& Henning 1999). 
Robinson (1995) suggested that the profiles of BLR emission of NLS1s 
are generally different
from those of BLS1; i.e., a normal broad-line profile has a more
dominant core than a NLS1 profile [O2]. 
Since there are two kinds of
BLR (i.e., a disk-like and a jet-like BLR in a Seyfert galaxy),
it is expected that the relative contribution of these two 
components to the total flux may be different from Seyfert to Seyfert.
Therefore, it may not be surprising to observe various kinds of Balmer
emission-line profiles.
The anti-correlation between Fe {\sc ii} $\lambda$4570/H$\beta$ and
[O {\sc iii}]$\lambda$5007/H$\beta$ [O8] can be understood if 
the NLR has a significant amount of dust grains (Netzer \& Laor 1993)
because the NLS1s are viewed from lower viewing angles and thus the 
effect of extinction on the [O {\sc iii}] emission
could be more serious than that for the BLS1s.

Since the HI-BLR emission lines are thought to arise from 
the jet-like BLR, there can be little difference in the UV emission
line properties ([U1] and [U2]) between the NLS1s and the BLS1s.
Since the UV -- soft X-ray emitting regions in NLS1s
are moving toward us at relativistic velocities, the UV part
of the continuum emission
is blueshifted to the soft X-ray regime and thus the UV luminosities
of NLS1s are expected to be less luminous than those of BLS1s 
systematically [U3].
This effect may also contribute to the softer hard X-ray spectra 
observed in NLS1s [X3].

Walter \& Fink (1993) found the positive correlation between  
$\Gamma_{\rm soft}$ and the UV (1375 \AA)-to-X ray (2 keV) flux ratio 
for S1s and NLS1s follow this correlation [X8].
In our model the UV continuum is mainly supplied by the 
optically-thick accretion disk. In order to explain this correlation,
it will be necessary to construct a more general model including the
accretion disk emission.  

It is worth noting the radio continuum image of one of NLS1s,
Mrk 766 (Ulvestad et al. 1995). Since this NLS1 is a very nearby object,
this is fortunately resolved spatially in the 3.6 cm and
6 cm radio continuum maps. The linear size is 60 pc
although the shape is slightly elongated in position angle of
$12^\circ \pm 5^\circ$. A typical linear size of the radio jets
in BLS1s is 350 pc (Ulvestad \& Wilson 1989).
If the small radio size of Mrk 766 is attributed to the inclination effect,
we obtain a viewing angle toward the jet is $i_{\rm view}$ =
sin$^{-1}$ ($60/350$) $\approx 10^\circ$, being consistent with those
derived from the kinematical and statistical considerations.
Since most NLS1s are not resolved in the observations of Ulvestad et al.
(1995), the pole-on view model appears consistent with the radio 
observations in this respect [R2].
However, the following observations are not interpreted easily;
two NLS1s (Mrk 766 and Mrk 1126)
have radio major axes perpendicular to the optical polarization
while the remaining one (Mrk 957) has a radio major axis
parallel to the optical polarization [R3].
It has been shown that the optical polarization position angles
tend to align with the radio-jet structure in type 1 AGN while
be perpendicular to the radio-jet structure in type 2 AGN
(Antonucci 1983, 1984, 1992). If NLS1s belong to a subclass of S1s,
we would obtain that all the above three NLS1s have radio major axes
parallel to the optical polarization.

As mentioned in section 5.2 briefly,
accretion disks in AGNs have been probed by
the very broad Fe K$\alpha$ emission (Tanaka et al. 1995;
Fabian et al. 1995; Mushotzky et al. 1995; Iwasawa et al. 1996;
Nandra et al. 1997; Reynolds 1997).
Nandra et al. (1997) presented a systematic analysis of
the hard X-ray spectra of 18 BLS1s.
They fitted the Fe K$\alpha$ emission profiles and derived the most probable
inclination angles; $\overline{i}_{\rm view} \simeq 29^\circ \pm 3^\circ$.
Although these estimates are subject to a number of fitting parameters,
it is interesting to note that most of the BLS1s are viewed from such
intermediate viewing angles.

\subsection{Comments on the GBHC Model}

It has been often claimed that the soft excess emission in NLS1s
is similar to the observational property of the high state of
Galactic black hole candidates (GBHCs) like Cyg X-1
(Tanaka 1990; Pounds et al. 1995; Hayashida 1997).
These GBHCs are known to show a dramatic change in spectral shape
from a normal AGN-like power law to a state where the soft
X-ray emission dominates, perhaps being attributed to the thermal 
emission from the accretion disk.
Although the power-law spectrum can still be seen in this high state,
it is often observed to be steeper than before as a result of
the increased electron cooling from the enhanced soft X-ray photons.
This spectral change may be caused by an increase in the accretion rate;
the accretion rates in the GBHCs are thought to be close to the
Eddington limit.
However, as discussed in section 7.3, the soft X-ray luminosities of
NLS1s are extrinsically more enhanced by the kinematical Doppler factor
than those of BLS1s. Therefore, it is not necessary to interpret that
NLS1s have higher gas accretion rates than BLS1s.

BBF96 also noted the following difficulty for the GBHC model;
GBHCs tend to be less variable while in their ultra soft states,
while NLS1s do not show reduced variability. 

\section{A VIEWING-ANGLE-DEPENDENT UNIFIED MODEL FOR SEYFERT GALAXIES}

\subsection{The Critical Viewing Angle to NLS1s}

As discussed in section 5.3, the pole-on view model
implies that the observed narrow FWHMs of 
NLS1s are attributed to smaller viewing angles toward the BLR.
A critical viewing angle toward NLS1s is $i_{\rm cr, NLS1} \simeq$
sin$^{-1}$ (1000 km s$^{-1}$/12000 km s$^{-1}$) $\simeq 4\fdg8$
or  $i_{\rm cr, NLS1} \simeq 9\fdg6$
if we adopt the maximum value of
FWHM(BLR) = 2000 km s$^{-1}$ for NLS1s (Vaughan et al. 1999).
Since S1s with FWHM(BLR) $\simeq$ 2000 km s$^{-1}$ are often
classified as NLS1s, the latter viewing angle seems a more appropriate 
estimate. Therefore, we adopt $i_{\rm cr, NLS1} \simeq 10^\circ$.

\subsection{The Critical Viewing Angle to BLS1s}

After introducing the dusty torus model for Seyfert nuclei
(Antonucci \& Miller 1985), many arguments have been made
to estimate the critical viewing angle between S1s and S2s
(Osterbrock \& Shaw 1988; Salzer 1989; Miller \& Goodrich 1990;
Huchra \& Burg 1992; Pogge 1989; Wilson \& Tsvetanov 1994;
Schmitt \& Kinney 1996; Murayama et al. 2000).

a) The statistical method:
The critical viewing angle $i_{\rm cr, BLS1}$ can be 
estimated from the number statistics
of S1s and S2s if we observe Seyfert nuclei from random orientations on
the statistical ground,

\begin{equation}
{N_{\rm S1} \over {N_{\rm S1} + N_{\rm S2}}} = 1 - \cos ~ i_{\rm cr, S1} \:
{\rm (stat)},
\end{equation}
where $N_{\rm S1}$ and $N_{\rm S2}$ are the observed numbers of S1s and S2s,
respectively (Miller \& Goodrich 1990).
The three different optical surveys of
Seyfert galaxies give the following critical angles;
$i_{\rm cr, BLS1} ({\rm stat})\simeq 27^\circ$ (Osterbrock \& Shaw 1988),
34$^\circ$ (Salzer 1989), and 46$^\circ$ (Huchra \& Burg 1992).

b) The opening angle of NLR:
The NLR of S2s often exhibits conical morphologies, which are
due to shadowing of the nuclear ionizing continuum by the torus. The
observed semi-opening angle of the cone $\theta_{\rm open} ({\rm NLR})$ is
thereby equal to the semi-opening angle of the torus $\theta_{\rm open}$
as noted in section 6.2 briefly.
This angle can be close to the critical viewing angle to BLS1s.
The statistical results from observations of conical
NLRs give the following opening angle;
$\theta_{\rm open} ({\rm NLR}) = 26^\circ \pm 11^\circ$  (Pogge 1989),
$32^\circ \pm 8^\circ$ (Wilson \& Tsvetanov 1994), and 
$29^\circ \pm 9^\circ$ (Schmitt \& Kinney 1996).

c) The mid-infrared diagnostic:
Since the $L$-band (3.5 $\mu$m)-to-IRAS 25 $\mu$m flux ratio is 
sensitive to the orientation of dusty tori, this ratio can be used to
estimate the viewing angle to the dusty tori although
dusty torus models are necessary to perform this analysis
(Taniguchi et al. 1997; Murayama et al. 2000).
Murayama et al. (2000) found that the critical viewing angle toward
BLS1s is about $45^\circ$.

In summary, the critical viewing angle to BLS1s lies in a range
between $30^\circ$ and $45^\circ$. Since the opening angle of the NLR
provides only a direct estimate, we adopt $i_{\rm cr, BLS1} \simeq 30^\circ$.

\subsection{The Critical Viewing Angle to S2s with the Hidden BLR}

S2s are also classified into the following two classes; 1)
S2s with the hidden BLR (hereafter S2$^+$), and 2)
S2s without the hidden BLR (hereafter S2$^+$)
(e.g., Miller \& Goodrich 1990; Tran 1995). 
Now let us assume that
the S2$^+$s are viewed at an angle intermediate
between S1s and S2$^-$s (Heisler, Lumsden, \& Bailey 1997).
We can estimate a critical viewing angle which distinguishes S2$^+$ from
S2$^-$ ($i_{\rm cr, S2^+}$),

\begin{equation}
{{N_{\rm S1} + N_{\rm S2^+}} \over {N_{\rm S1} + N_{\rm S2}}} =
1 - {\rm cos} ~ i_{\rm cr, S2^+}.
\end{equation}
Using this relation together with 
the observed number ratio between S2$^+$ and S2$^-$,
$N_{\rm S2^+}/N_{\rm S2^-}$ = 10/40 = 0.25,
Taniguchi \& Anabuki (1999) obtained
$i_{\rm cr, S2^+} \approx 50^\circ$. 
This means that Seyfert galaxies viewed from $50^\circ \lesssim
i_{\rm view} \leq 90^\circ$ are identified as S2$^-$s.

\subsection{Summary}

The above arguments lead us to propose a viewing-angle-dependent
unified model for Seyfert nuclei;
1) $0^\circ \leq i_{\rm view} \lesssim 10^\circ$ for NLS1s,
2) $10^\circ \lesssim i_{\rm view} \lesssim 30^\circ$ for BLS1s,
3) $30^\circ \lesssim i_{\rm view} \lesssim 50^\circ$ for S2$^+$s, and
4) $50^\circ \lesssim i_{\rm view} \leq 90^\circ$ for S2$^-$s.
This is schematically illustrated in Figure 7.

\begin{figure*}
\epsscale{1.0}
\plotone{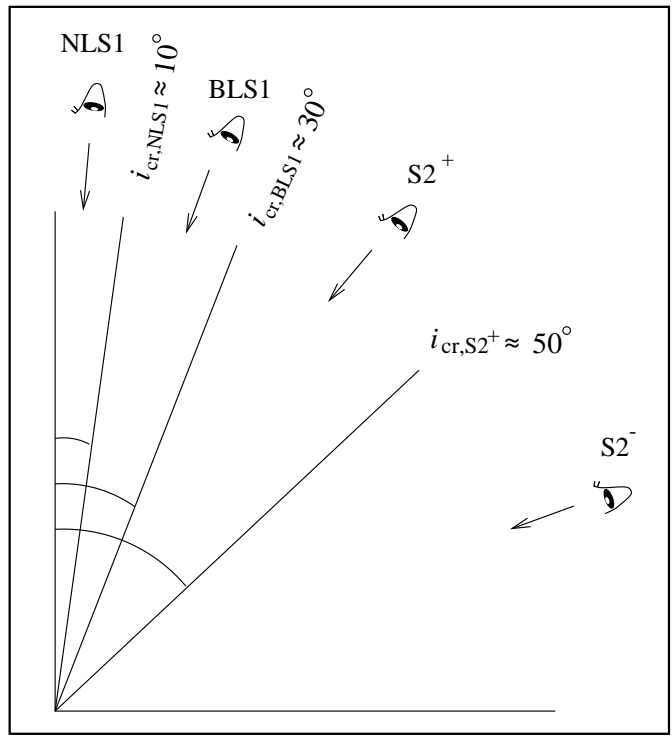}
\caption{
A viewing-angle-dependent unified model for Seyfert galaxies
(NLS1, BLS1, S2$^+$, and S2$^-$).
\label{fig7}
}
\end{figure*}

Finally, it seems important to note that the NLS1 phenomenon
is also observed in some radio-quiet quasars
(Laor et al. 1994, 1997; Fiore et al. 1998; Xu et al. 1999).
Therefore, it is strongly suggested that the class of NLS1s 
is the radio-quiet equivalent of the class of Blazers
in radio-loud AGN (see for unified models for radio-loud
AGN, Urry \& Padovani 1995).

\vspace{0.5cm}

We would like to thank Kazushi Iwasawa, Neil Brandt, Hisamitsu Awaki,
and Kiyoshi Hayashida for useful discussion on the X-ray properties
of Seyfert nuclei and Hideaki Mouri and Deborah Dultzin-Hacyan
for useful discussion on AGN.
We also thank our colleagues, Shingo Nishiura, 
Hiroshi Sudou, and Naohisa Anabuki for useful discussion 
during the course of this work.
TM is a JSPS fellow. This work was financially supported 
in part by Grant-in-Aids for the Scientific
Research (Nos. 10044052, and 10304013) of the Japanese Ministry of
Education, Culture, Sports, and Science.


\end{document}